\documentclass[fleqn,usenatbib]{mnras}
\usepackage{newtxtext,newtxmath}
\usepackage[T1]{fontenc}

\DeclareRobustCommand{\VAN}[3]{#2}
\let\VANthebibliography\thebibliography
\def\thebibliography{\DeclareRobustCommand{\VAN}[3]{##3}\VANthebibliography}

\usepackage{graphicx}	
\usepackage{amsmath}	
\usepackage[dvipsnames]{xcolor}
\usepackage{enumitem}
\usepackage{url}
\usepackage{braket}
\usepackage{threeparttable}
\usepackage{hyperref}
\usepackage{multirow}
\usepackage{tabularx}
\usepackage{tabulary}

\def\myerr[#1]{{\color{red} #1}}
\def\myemph[#1]{{\color{blue} #1}}
\def\hjmo[#1]{{\color{green} #1}}
\def\myrevise[#1]{{\bf #1}}
\def\myrevises[#1]{{\color{red} #1}}
\def\term[#1]{{\bf \ttfamily #1}}
\def\B[#1]{{\bf #1}}

\setlist[itemize,1]{label=$\bullet$}
\setlist[itemize,2]{label=$\bullet$}
\setlist[itemize,3]{label=$\bullet$}
\setlist[itemize,4]{label=$\bullet$}
\setlist[itemize]{leftmargin=*}
\def\bit{\begin{itemize}[topsep=0em,parsep=0em,itemsep=0em,partopsep=0em,leftmargin=1.0em]}
\def\bitt{\begin{itemize}[topsep=0em,parsep=0em,itemsep=0em,partopsep=0em,leftmargin=3.0em]}
\def\eit{\end{itemize}}

\def\beq{\begin{equation}}
\def\eeq{\end{equation}}
\def\bey{\begin{eqnarray}}
\def\eey{\end{eqnarray}}

\def\bfrm[#1]{\mathrm{{\bf#1}}}

\def\gs{\mathrel{\raise1.16pt\hbox{$>$}\kern-7.0pt
		\lower3.06pt\hbox{{$\scriptstyle \sim$}}}}
\def\ls{\mathrel{\raise1.16pt\hbox{$<$}\kern-7.0pt
		\lower3.06pt\hbox{{$\scriptstyle \sim$}}}}
\def\gtsima{\, {\buildrel > \over \sim} \,}
\def\ltsima{\, {\buildrel < \over \sim} \,}
\def\prosima{\, {\buildrel \propto \over \sim} \,}
\def\gsim{\lower.5ex\hbox{\gtsima}}
\def\lsim{\lower.5ex\hbox{\ltsima}}
\def\simgt{\lower.5ex\hbox{\gtsima}}
\def\simlt{\lower.5ex\hbox{\ltsima}}
\def\simpr{\lower.5ex\hbox{\prosima}}
\def\la{\lsim}
\def\ga{\gsim}

\def\Mpc{\,{\rm Mpc}}

\def\gyri{\, h\,{\rm Gyr}^{-1}}

\def\Msun{{\rm M_\odot}}
\def\msun{\, h^{-1}{\rm M_\odot}}

\def\Rvir{R_{\rm vir}}
\def\Vvir{V_{\rm vir}}

\def\vmax{v_{\rm max}}
\def\Mhalo{M_{\rm halo}}

\def\Mstar{M_{*}}
\def\MstarInt{M_{\rm *,\,int}}

\def\halospin{\lambda_{\rm s}}
\def\haloshape{q_{\rm axis}}

\def\halopc[#1]{{\rm PC}_{\rm MAH, #1}}
\def\halovmaxpcs{{\rm PC}_{\rm vmax}}
\def\halovmaxpc[#1]{{\rm PC}_{\rm vmax, #1}}
\def\zfrom[#1]{z_{\rm #1}}

\def\haloAccrRate{ \langle \dot{M}_{\rm halo} \rangle }
\def\hhalo{ \mathrm{\bf h}_{\rm halo} }
\def\xhalo{ \mathrm{\bf x}_{\rm halo} }
\def\hstar{ \mathrm{\bf h}_{\rm *} }
\def\xstar{ \mathrm{\bf x}_{\rm *} }

\def\ssfr{{\rm sSFR}}
\def\ssfrRes{\Delta\,\log\,{\rm sSFR}}

\def\mstarpc[#1]{{\rm PC}_{\rm M_*, #1}}

\title[]{
How to empirically model star formation in dark matter halos:  
I. Inferences about central galaxies from numerical simulations}

\author[Yangyao Chen et al.]{
	Yangyao Chen,$^{1,2}$\thanks{E-mail: yangyaochen.astro@foxmail.com} 
	H.J. Mo, $^{2}$   
	Cheng Li$^{1}$  
	and Kai Wang$^{1,2}$ 
	\\
	$^{1}$Department of Astronomy, Tsinghua University, Beijing 100084, China\\
	$^{2}$Department of Astronomy, University of Massachusetts, Amherst, MA 01003-9305, USA\\
}

\date{Accepted XXX. Received YYY; in original form ZZZ}

\pubyear{2020}

\begin{document}
\label{firstpage}
\pagerange{\pageref{firstpage}--\pageref{lastpage}}
\maketitle

\begin{abstract}
	We use TNG and EAGLE hydrodynamic simulations to investigate the 
	central galaxy - dark matter halo relations that are needed for 
	a halo-based empirical model of star formation in galaxies.
	Using a linear dimension reduction algorithm and a model ensemble 
	method, we find that for both star-forming and quenched galaxies, 
	the star formation history (SFH) is tightly related to the halo 
	mass assembly history (MAH). 
	The quenching of a low-mass galaxy is mainly due to the 
	infall-ejection process related to a nearby massive halo, 
	while the quenching of a high-mass galaxy is closely related to 
	the formation of a massive progenitor in its host halo. The 
	classification of star-forming and quenched populations 
	based solely on halo properties contains contamination 
	produced by sample imbalance and overlapping distributions of 
	the two populations. Guided by the results from hydrodynamic 
	simulations, we build an empirical model to predict the SFH 
	of central galaxies based on the MAH of their host halos, and 
	we model the star-forming and quenched populations separately. 
	Our model is based on the idea of adopting star formation 
	templates from hydrodynamic simulations to reduce model 
	complexity. We use various tests to demonstrate that the 
	model can recover star formation histories of 
	individual galaxies, and can statistically reproduce the 
	galaxy bimodal distribution, stellar mass - halo mass 
	and star formation rate - halo mass relations from low to 
	high redshift, and assembly bias. Our study provides a 
	framework of using hydrodynamic simulations to discover, 
	and to motivate the use of, key ingredients to model 
	galaxy formation using halo properties. 
\end{abstract}
\begin{keywords}
	halos -- galaxies -- formation -- stellar content -- hydrodynamic 
\end{keywords}

\section{Introduction}
\label{sec:intro}

In the $\Lambda$CDM cosmology, galaxies are luminous objects that form and evolve in the gravitational potential wells 
of their dark matter halos in the cosmic density field. A key step to understand how galaxies 
form and evolve is, therefore, to understand how galaxies are related 
to dark matter halos \citep[see][and references therein]{MoBoschWhite:2010:GFE, WechslerRisaH:2020:AnnRiev:Galaxy-halo-connection}.
Because the dark matter is invisible, direct observation is 
inaccessible. Meanwhile, numerical simulations based on first principles
have also limitations, because of the use of  sub-grid physical processes
that are not resolved. Because of these difficulties, 
a variety of other methods, generally referred to as 
empirical models, have been developed to link galaxies with 
dark matter halos. The details of these models, such as the model 
architectures and model parameters, can be constrained by 
observations, such as the galaxy stellar mass functions (GSMFs), 
two-point correlation functions (2PCFs), and so on. 
Examples of such models include (sub)halo abundance matching \citep{
	MoHJ:MaoS:WhiteSDM:1999:LBG:AM, 
	Vale2004, 
	GuoQi:2010:SHAM}, 
clustering matching \citep{GuoHong_ZhengZheng_2016_SHAM_SCAM}, 
age matching \citep{
	HearinAP:2013:SubhaloAgeMatching,
	HearinAP:2014:Age-matching-2,
	MengJiacheng:2020:High-z-mock-and-measurement}, 
halo occupation distribution \citep{JingYP:1998:HOD, BerlindAA:2002:HOD},
conditional luminosity function \citep{YangXiaohu:2003:CLF_M2LRatioTurnOver}, 
conditional color-magnitude diagrams \citep{XuHJ:2018:CCMD},
and those based on star formation histories  
\citep{
	LuYu_MoHoujun_2011_Bayes_SAM,
	LuYu_MoHoujun_2014_SAM_Bayes,
	LuZhankui:2014:EmpiricalModel, 
	LuZhankui:2015:EmpiricalModel, 
	MosterBenjaminP_2018_EmpiricalModel, 
	BehrooziPeter:2019:UniverseMachine, 
	MosterBP:2020:GalaxyNet}. 
Although these empirical models are 
able to reproduce a large set of observations, 
some basic questions remain unresolved. 

First, when modeling the galaxy-halo connection, it is not totally 
clear which halo quantities are the best to use as features 
to make the link to galaxies, and which set of galaxy quantities 
is the best in constraining the link. To reproduce the 
primary galaxy properties, such as stellar mass and luminosity, 
it is likely that the basic properties of halos, such as 
halo mass and peak circular velocity, are the main features to 
use \citep[see, e.g.,][for an extensive study]{ReddickRM:WechslterRH:2013:SHAM:VmaxBest}. However, when higher order galaxy properties are concerned,  
a systematic approach is yet to be found to identify the best 
set of halo features for the purpose. 
\cite{MosterBP:2020:GalaxyNet} showed an example of using random 
forest regressor to find the important halo properties 
related to galaxy stellar mass and star formation rate (SFR).
This motivates, but has not been used in, the construction of a 
deep and dense model. 

Second, the total sets of galaxy and halo properties, which 
are high-dimensional, are too complex to be useful, 
because it is both difficult to incorporate them into models 
and to interpret their roles in model predictions. 
For example, the details of the formation histories of individual 
halos are complex, so are the star formation and merger histories 
of individual galaxies. Yet, it is necessary to include 
them in the modeling, as they carry important information 
connected to the current state of an object, such 
as halo concentration 
\citep[e.g.,][]{
	NavarroJ_FrenkC_WhiteS_1997_NFWProfile,
	JingYP:2000:HaloConcenScatterAndOnMAH,
	Wechsler:2002:HaloConcetrationAndMAHFitting,	
	ZhaoDH:MoHJ:2003MN:ZMJBModel:HaloConcentration,
	ZhaoDH:MoHJ:2003ApJ:NBodyValidationZMJBModel:HaloConcentration,
	MacCioAV:2008:HaloStructVsCosmology,
	ZhaoDonghai_2009_HaloConcenAnalytical,
	JeasonDanielAkila:2011:PCA:HaloProperties}, halo bias
\citep[e.g.,][]{MoHJ:WhiteSDM:1996:HaloBias, 
	ShethRK:MoHJ:TormanG:2001:EllipseBias,
	GaoL:SpringelV:2005:AssemblyBias,  WechslerRH:ZentnerAR:2006:AssemblyBiasAndOnConcenOrOccupation,
	GaoL:2007:HaloAssemblyBiasOnSpinOrSub,  
	JingYP:SutoY:MoHJ:2007:ApJ:ClusteringOnZform:Andc, BettPhilip:2007:HaloSpinShapeInMillSim:BiasOnSpin, HahnOliver:2007:HaloSpinShapeMAHOnCosmicWebClassif, 
	LiYun_MoHoujun_2008_HaloFormationTimesDef,
	FaltenbacherA:2010:HaloAssemblyOnManyProp, 
	WangHuiyuan_MoHoujun_2011_HaloDependOnEnv}, 
galaxy color or SFR \citep[e.g.,][]{
	ShiJingjing:2020:HaloMAHOnSatelliteProp,
	HearinAP:2013:SubhaloAgeMatching, HearinAP:2014:Age-matching-2,
	LimSH:2016:Halo-assembly-on-galaxies,
	MengJiacheng:2020:High-z-mock-and-measurement,
	WangHuiyuan2017_ELUCID4}, 
and galaxy structure \citep[e.g.,][]{
	KauffmannG:2003:SFH-struct-mass,
	ShenShiyin:2003:mass-size,
	BernardiM:2007:BCG-lumin-size-veldisp-formation, 
	GaoYing:2020:Compact-core-as-high-z-descendant,
	YoonYongmin:2020:Fundamental-plane-on-age-structure}.  
Attempts have been made to use various formation redshifts to describe 
halo assembly histories \citep[see, e.g.,][]{NavarroJ_FrenkC_WhiteS_1997_NFWProfile,vandenBosch:2002:HaloUniversalMAH,LiYun_MoHoujun_2008_HaloFormationTimesDef,ZhaoDonghai_2009_HaloConcenAnalytical,JeasonDanielAkila:2011:PCA:HaloProperties,WangHuiyuan_MoHoujun_2011_HaloDependOnEnv,ShiJJ2018_HaloBimodalFormation}. However, as shown in \cite{ChenYangyao:2020:Halo-structure}, 
the information provided by these formation times is incomplete, and 
there is strong degeneracy among them. 

Third, because of the complexity in the galaxy-halo connection, 
it is important to know how we construct a model that is 
general but can still felicitate clear physical interpretations 
of the results it produces. The ``performance-interpretation'' 
trade-off is a common problem in model construction. 
For example, the empirical model of \cite{LuZhankui:2014:EmpiricalModel} 
used a physically motivated relation between SFR and halo mass and 
redshift, while \cite{BehrooziPeter:2019:UniverseMachine} 
used the growth of peak circular velocity to rank the SFR.  
In both models, the physical meaning of the galaxy-halo 
connection is clear, but both may have missed other potentially 
important factors as well as nuanced processes that can cause 
uncertainties in the relations. 
In contrast, an empirical model based on densely connected 
neural networks, such as the one developed by
\cite{MosterBenjaminP_2018_EmpiricalModel}, usually 
uses multiple hidden layers to get a good representation 
of the halo properties and regresses them on stellar 
mass and SFR. The model is accurate, as long as there are 
sufficient constraints from observations, but 
the representation of halos is in a ``black-box'', 
making it difficult to interpret the results. 
Some other empirical models invoke multiple ingredients, 
for example, by separating central galaxies from satellites 
and/or star-forming from quenched galaxies,  
and use observations to constrain the joint distribution 
of model parameters \cite[e.g.,][]{LuZhankui:2014:EmpiricalModel, LuZhankui:2015:EmpiricalModel, 
MosterBenjaminP_2018_EmpiricalModel, 
BehrooziPeter:2019:UniverseMachine}. 
Even in such models, it is still challenging 
to show which ingredients dominate the prediction error, 
and whether the discrepancy with observations 
owes to the incapability of the model or to the incompleteness 
of the observation constraints.

In this paper, we carry our a systematic investigation of the 
ingredients that are needed to construct a powerful 
empirical model of galaxy formation based on dark matter halos.
We use inferences from hydrodynamic simulations to
motivate a potentially useful architecture to build
such an empirical model. To address the first 
problem described above, we adopt 
a model ensemble algorithm, the gradient boosted decision trees 
(GBDT, see Appendix~\ref{app:GBDT}), which can be used 
not only to capture complicated patterns between variables and 
to keep a good balance between bias and variance, 
but also to identify the most important variables that explain 
model predictions. We address the second problem 
by using a linear dimension reduction algorithm, 
the principal component analysis (PCA, see Appendix~\ref{app:PCA}), which can 
effectively reduce the dimension of the halo assembly 
history and the galaxy star formation history
and yet retain large amounts of information 
of the histories for the empirical modeling. 
Finally, we address the third problem  by building a 
deep model that incorporates components of both dimension 
reduction and ensemble regressor and classifier. Each component in the model 
is motivated physically and can be optimized separately. 
This approach makes the model capable of dealing 
with complex patterns in parameter space, and yet   
transparent to interpret. The ensemble regressor and linear 
dimension reduction method has already been used to study 
the relationship among halo properties in 
\cite{ChenYangyao:2020:Halo-structure}. 
Here we extend it to studying the galaxy-halo connection.
As the first in a series, this paper focuses on 
central galaxies in dark matter halos. We identify 
important ingredients that should be included in an 
empirical model, and demonstrate the limit 
such a model can reach in describing the stellar masses
and star formation histories of individual galaxies.  
Our model is built on the inferences from two 
hydrodynamic simulations, the Illustris-TNG 
\citep[e.g.,][]{NelsonD:SpringelV:2019:TNGDataRelease} and EAGLE \citep[e.g.,][]{EAGLE:2017:Data-release-particle}. 

This paper is organized as follows. In \S\ref{sec:data}, 
we describe the simulation data we use, and define the halo 
properties and samples used in our analysis.
In \S\ref{sec:galaxy-halo-conn}, we use both the GBDT and PCA 
to study the relations of galaxy stellar mass and SFR 
with halo properties for both star-forming galaxies and 
quenched galaxies. We also identify halo properties 
that cause a galaxy to quench. In \S\ref{sec:model}, 
we build an empirical model that predicts the SFH of 
galaxies in dark matter halos, testing its performance 
in several steps. We summarize and discuss our results in 
\S\ref{sec:summary}.

\begin{figure}
	\includegraphics[width=\columnwidth]{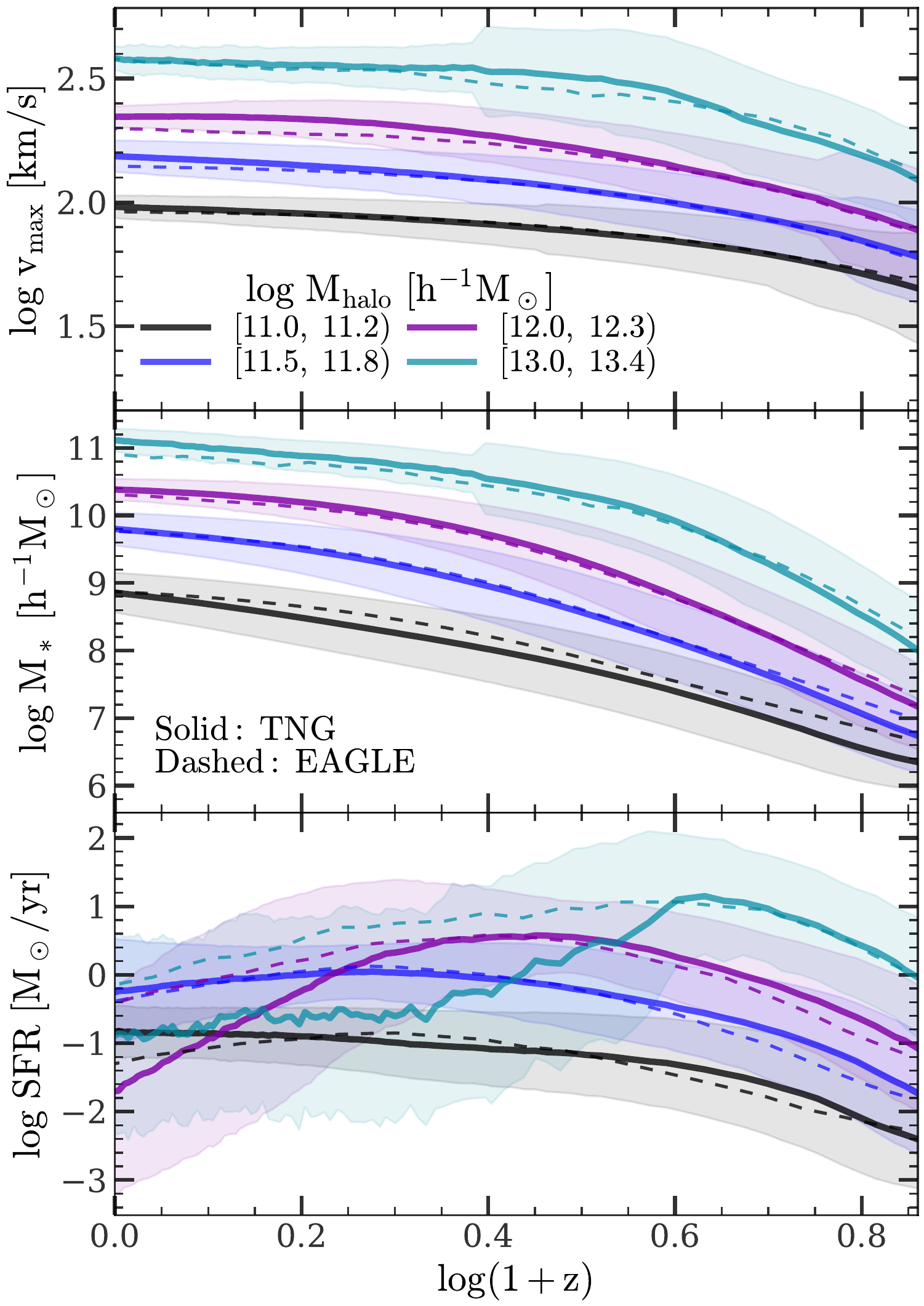}
	\caption{
		Halo mass assembly histories (MAHs) of central galaxies, characterized by $\vmax$ ({\bf top}), $\Mstar$ ({\bf middle}) and SFR ({\bf bottom}) as functions of $\log(1+z)$. In each panel, results are shown for four different ranges of final ($z=0$)
		halo masses, and separately for TNG and EAGLE in 
		solid and dashed lines, respectively. The 1-$\sigma$ scatter is shown 
		for the TNG simulation only.
	}
	\label{fig:histories}
\end{figure}

\begin{figure}
	\includegraphics[width=\columnwidth]{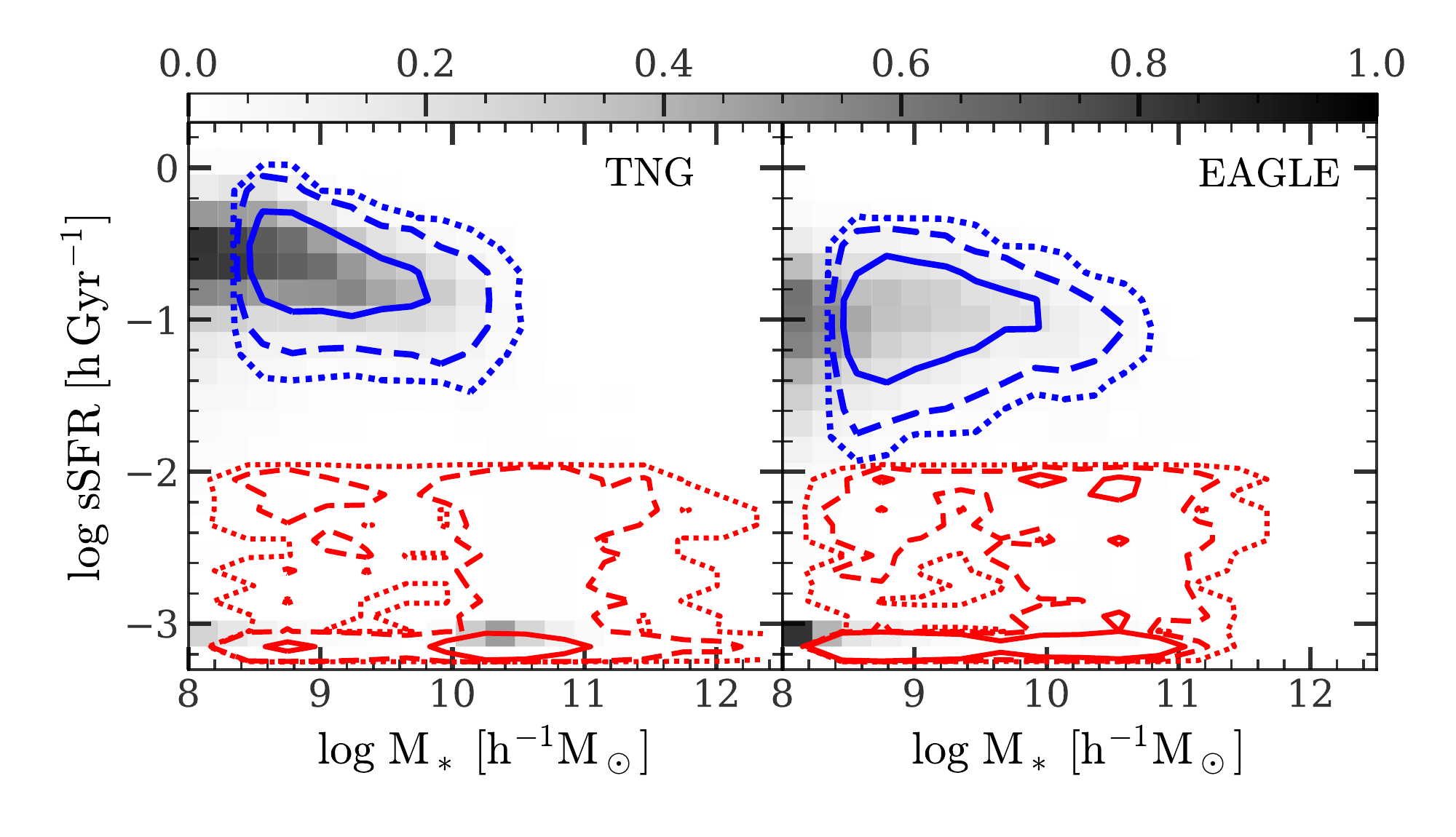}
	\caption{Distribution of central galaxies at $z=0$ in the plane of
          $\log\,\ssfr$ versus $\log\,\Mstar$, for TNG ({\bf left}) and
          EAGLE ({\bf right}). The {\color{Gray} \bf gray} shade
          shows the normalized distribution of the full sample. 
          Galaxies above $10^{8.5} \msun$ are divided into the star-forming
          sample, $\rm S'_{SF}$ ({\color{Blue}\bf blue} contours), 
          and the quenched sample, $\rm S_Q$ ({\color{Red}\bf red} contours).
          Solid, dashed and dotted contours enclose 1, 2 and 3$-\sigma$ regions,
          respectively. Galaxies with sSFR below $10^{-3}\gyri$ are stacked
          at the bottom of the panel. See \S\ref{ssec:sample} and
          Table~\ref{tab:def-samples} for the definitions of different samples.
	}
	\label{fig:sample-separation}
\end{figure}

\section{The Data}
\label{sec:data}

\subsection{The Illustris-TNG and EAGLE simulations}
\label{ssec:sim}

The Illustris-TNG simulation \citep{NelsonD:SpringelV:2019:TNGDataRelease, Pillepich:2018:TNG-first-result-stellar, SpringelV:2018:TNG-first-results-clustering, NelsonDylan:2018:TNG-first-results-color, NaimanJillP:2018:TNG-first-results-chemical,Marinacci:2018:TNG-first-results-magnetic} is a suite of cosmological, hydrodynamical simulations implemented with 
the moving-mesh code {\tt Arepo} \citep{SpringelV:2010:Arepo}. The cosmological parameters are taken from the Planck 2015 results \citep{Planck15:Cosmo-param}: Hubble constant $H_0=100 h\, {\rm km\,s^{-1}\,Mpc^{-1}}$ with $h=0.6774$, cosmological constant $\Omega_{\rm \Lambda,0}=0.6911$, matter density $\Omega_{\rm M,0}=0.3089$, baryon density $\Omega_{\rm B,0}=0.0486$, and initial power spectrum 
with normalization $\sigma_8=0.8159$ and index $n_{\rm s}=0.9667$.
The simulated physical processes for galaxy formation include 
gas cooling, star formation, stellar feedback, metal enrichment, 
black hole feedback, and so on. The details can be found in the two method 
papers, \cite{WeinbergRainer:2017:TNG-AGN-model} and  \cite{PillepichA:2018:TNG-physical-model}. 
A total of 100 snapshots are saved for each of the simulation runs.
Halos are identified with the friends-of-friends (FoF) 
algorithm \citep{DavisM1985_FOF} and subhalos are identified with 
the SUBFIND algorithm \citep{SpringelVolker_2001_SUBFIND, DolagK:2009:SUBFIND}. 
Subhalo merger trees are constructed by the SubLink algorithm \citep{RodriguezGomezV:2015:Illustris:GalMergeTree:MergeRate}. To achieve a balance between sample size and resolution, we choose to use the TNG100-1 run 
(thereafter TNG), which has a box size with co-moving volume 
$(106.5 \Mpc)^3$, $2\times 1820^3$ resolution units, 
a target baryon mass resolution of $1.4\times 10^6 \Msun$, and 
dark matter particle mass of $7.5\times 10^6 \Msun$.

The EAGLE project \citep{SchayeJoop2014:EAGLE-intro-galaxy, CrainRobertA:2015:EAGLE-intro-calibration, McAlpine:2016:Eagle-release-halo-galaxy, EAGLE:2017:Data-release-particle} consists of a suite of cosmological hydrodynamic simulations performed with the GADGET-3 tree-SPH code, which is an extension of the GADGET-2 code \citep{Springel2005}. The cosmological parameters are taken from the Planck 2013 results \citep{AdePAR:2014:Planck-2013-overview-science}: $h=0.6777$, $\Omega_{\rm \Lambda,0}=0.693$, $\Omega_{\rm M,0}=0.307$, $\Omega_{\rm B,0}=0.04825$, $\sigma_8=0.8288$, and $n_{\rm s}=0.9611$. The subgrid processes simulated 
include gas cooling and heating, star formation, stellar evolution, metal enrichment, stellar feedback, and black hole feedback. A total of 29 snapshots are 
saved for each of the runs. Halos and subhalos are also identified with FoF 
and SUBFIND algorithms. Subhalo merger trees are constructed 
using the D-TREES algorithm \citep{JiangL:2014:DTREES}. 
To achieve a balance between sample size and resolution, we choose to use 
EAGLE Ref-L0100N1504 (thereafter EAGLE), which has a box size 
with a co-moving volume of $(100 \Mpc)^3$,  a total number of particles of  
$2\times 1504^3$, initial baryonic particle mass of 
$1.81\times 10^6\Msun$, and dark matter particle mass of 
$9.70\times 10^6\Msun$.

The output galaxy and halo catalogs in TNG and EAGLE present a variety of quantities, 
such as halo mass, galaxy stellar mass, and SFR. 
Using the method described by \cite{RodriguezGomezV:2015:Illustris:GalMergeTree:MergeRate},
we construct merger trees for FoF halos from the subhalo merger trees, 
Whenever a needed galaxy or halo property is not included in 
the public catalog, we calculate it using the particle/cell data.
The FoF halo and subhalo properties used in our analysis are listed below.
\bit 
\item $\Mhalo$: ``top-hat'' mass of the FoF halo within a radius where the overdensity is that given by the spherical collapse model \citep{Bryan1998}.

\item $M_{\rm subs}$: for a central subhalo (defined as the most massive subhalo in TNG, and the most bound subhalo in EAGLE), it is the total  
mass bounded to all subhalos in an FoF halo; for a satellite 
subhalo, it is the mass bounded to the subhalo itself.

\item $\vmax$: peak circular velocity of a subhalo, $\sqrt{G M(<r)/r}$, where $M(<r)$ is the total mass within a radius $r$.

\item $z_{\rm lmm}$: the redshift of last major merger 
of an FoF halo, where a major merger is defined as a merger event 
with the mass ratio between the small and large progenitors  
larger than one-third.

\item $z_{\rm infall}$: last in-fall redshift of an FoF halo, defined as 
the lowest redshift at which a progenitor of the central subhalo of 
the subhalo merger tree is not the most massive subhalo 
in the hosting FOF halo.

\item $z_{\rm mb,\,1/2}$: the highest redshift at which the main-branch 
progenitor of an FoF halo in the FoF halo merger tree assembled 
half of its final mass $\Mhalo$.

\item  $z_{\rm mb,\,core}$: the highest redshift at which the 
main-branch progenitor of an FoF halo in the FoF halo merger tree 
reached a fixed mass $M_{\rm h,\, core}=10^{11.5}\msun$.

\item $c$: the concentration parameter of the Navarro-Frenk-White (NFW) profile \citep{NavarroJ_FrenkC_WhiteS_1997_NFWProfile} of an FoF halo.

\item  $\haloshape$: the shape parameter, $(a_2+a_3)/(2a_1)$, of an FoF halo, where $a_1\ge a_2\ge a_3$ are the lengths of the three axes of the 
inertia ellipsoid. Only particles within $2.5 r_{\rm s}$ are used, 
where $r_{\rm s}$ is the characteristic radius of the NFW profile.

\item $\halospin$: the dimensionless spin parameter of an FoF halo, defined as 
\begin{equation}
\halospin = \frac{\|\bfrm[j]\|}{\sqrt{2}\Mhalo \Rvir \Vvir},
\end{equation}
where $\bfrm[j]$, $\Rvir$ and $\Vvir$ are the total angular momentum, virial radius and virial velocity, respectively. Only particles within $2.5 r_{\rm s}$ are used.

\item $\haloAccrRate$: FoF halo accretion rate, defined as 
\begin{equation}
\haloAccrRate = \left\langle \frac{{\rm d}M_{\rm subs}}{{\rm d}t} \right\rangle _{\rm dyn} - 4\pi \Rvir^2 \rho(\Rvir) \left\langle \frac{{\rm d}\Rvir}{{\rm d}t} \right\rangle _{\rm dyn},
\end{equation}
where $\langle {\rm d}x/{\rm d}t \rangle _{\rm dyn} = [ x(t)-x(t-t_{\rm dyn}) ]/t_{\rm dyn}$ is the average growth rate of a quantity $x$. This rate 
is defined using the main branch of the subhalo merger tree 
rooted in the central subhalo, and the dynamical time is for
the FoF halo, $t_{\rm dyn}=\sqrt{\Rvir^3/(GM_{\rm subs})}$.

\item $d_{\rm ngb}$: distance of a halo to its nearest FoF halo 
whose $\Mhalo$ is larger than that of the halo in consideration.
\eit

Both $\Mhalo$ and $\vmax$ are provided by the TNG and EAGLE halo catalogs. 
We refer the reader to \cite{ChenYangyao:2020:Halo-structure}
for detailed definitions and physical meanings of 
$z_{\rm lmm}$, $z_{\rm mb,\,1/2}$, $z_{\rm mb,\,core}$, $c$, 
$\haloshape$ and $\halospin$. The details of $\haloAccrRate$ can be found in \cite{MosterBenjaminP_2018_EmpiricalModel}. 
We use the following quantity as our time (redshift) variable:
$\delta_{\rm c}(z)\equiv \delta_{\rm c,\,0}/D(z)$, 
where $\delta_{\rm c,\,0}=1.686$ is the critical overdensity 
for spherical collapse, and $D(z)$ is the linear growth factor at $z$. 
We use the transfer function given by \cite{Eisenstein1998}, and  
the linear growth factor given by \cite{Carroll1992a}. 

The galaxy properties used for our analysis are the following:
\bit 
\item $\Mstar$: the stellar mass of a galaxy, which is the sum of 
the masses of stellar particles within a radius that 
is two times the stellar half mass radius for TNG, or within 
30 physical $\rm kpc$ for EAGLE.

\item SFR: the star formation rate of a galaxy, defined as the sum 
of the SFR of gas cells/particles within the same radius as 
that used for $M_*$.

\item $\MstarInt$: the stellar mass that has ever formed in the history, i.e., $\sum_{n}{\rm SFR}_n \Delta t_n$, where ${\rm SFR}_n$ and $\Delta t_n$ are the SFR at the $n$th snapshot and the time interval spanned by this snapshot, respectively. So defined, $\MstarInt$ is different from $\Mstar$ in that the mass loss due to stellar evolution and mass change due to merger are not taken into account. 
However, if a merger event triggers a change in the in-situ SFR, 
its effect is indirectly contained in $\MstarInt$.

\item sSFR: the specific star formation rate, defined as ${\rm SFR}/M_*$.
\eit 

Due to the limited resolution of the simulations, the SFRs have large 
fluctuations among different snapshots. To make the results more 
stable, whenever necessary we smooth the data by averaging the SFRs 
in adjacent snapshots. We use five adjacent snapshots for the 
smoothing for TNG and two for EAGLE. The resulting 
SFR and sSFR are referred to as the smoothed SFR and sSFR, 
respectively.

The galaxy-halo connection is expected to depend not only on 
the current status of galaxies and halos, but also on their 
histories. We therefore define a number of ``history'' 
quantities to describe the formation histories of galaxies and halos. 
The halo assembly history (AH; or mass assembly history, MAH) of a subhalo is 
defined as the set of $\vmax$ values (a vector)
in the main branch of the subhalo merger tree 
rooted in the subhalo in question. Such a set is denoted as 
$\bfrm[v]_{\rm max}$\footnote{To avoid confusion, 
	we use ``$\log$'' to denote 10-based logarithm, 
	bold-roman characters to denote vectors.
	We use 1, 2 and 3-$\sigma$ regions to denote those covering $68\%$, $95\%$ and $99.7\%$ data points, respectively, in the space of any dimension.
	}, and has  
a dimension the same as the number of snapshots spanned by 
the merger tree. The galaxy star formation history (SFH) describes 
the amount of star formation in its history. As we are interested 
in both the SFR and the cumulative quantities, 
$\Mstar$ and $\MstarInt$, the SFH of a galaxy (or of a hosting subhalo) 
may refer to the set of values for SFR, or $\Mstar$, or $\MstarInt$, 
along the main branch of the subhalo merger tree, depending on the context. 
We denote the SFH described by these three quantities as 
$\bfrm[SFR]$, $\bfrm[M]_*$ and $\bfrm[M]_{\rm *,\,int}$, respectively, 
which are vectors with the same dimension as $\bfrm[v]_{\rm max}$. 
To avoid ambiguity, we refer $\bfrm[v]_{\rm vmax}$, $\bfrm[SFR]$, $\bfrm[M]_*$ and $\bfrm[M]_{\rm *,\,int}$ as the $\vmax$ history, the SFR history, 
the $\Mstar$ history and the $\MstarInt$ history, respectively.

All of the four history vectors are in the space of a too high 
dimension to be useful. Here we apply the same dimension 
reduction technique as used in \cite{ChenYangyao:2020:Halo-structure} 
to reduce the dimension of the history quantities. We provide a 
brief description of this method and its performance in 
Appendix~\ref{app:PCA}. After such dimension reduction, each 
of these histories becomes a set of principal components (PCs), 
which we denote as $\bfrm[PC]=({\rm PC_1},\,{\rm PC_2},\,...)$, 
with a subscript to distinguish different physical quantities.
The same technique was used in \cite{Chaves-MonteroJ:HearinA:2020:PCA-on-color-and-SFH}
to find the principal direction of galaxy distribution in the color space and to relate the 
principal color component to the SFH.

Fig.~\ref{fig:histories} shows the $\vmax$ history, the SFR history 
and the $\Mstar$ history for central subhalos of different masses 
obtained from TNG and EAGLE. Despite of the difference between 
the two simulations, some common patterns do exist. 
First, the histories of $\Mstar$ are very similar to those of $\vmax$,
both increasing with cosmic time, but the increase being 
slower at lower redshift. Second, the galaxy with a higher $\vmax$ 
also has a higher $\Mstar$ on average. Third, 
the SFR increases with time at high redshift but decreases 
at low redshift. This can also be seen from the fast-to-slow increase 
of $\Mstar$ with time and indicates that many of the galaxies 
become quenched at low redshift. All these suggest that 
the galaxy SFH is tightly correlated with halo assembly history, 
as we will quantify in the following sections.

\begin{center}
\begin{table*}
\caption{Galaxy samples used in this paper. Detailed definitions can be found in \S\ref{ssec:sample}. All of the samples are central galaxies selected by stellar mass and smoothed sSFR from TNG and EAGLE. }
\begin{tabularx}{\textwidth}{ c  X } 
	\hline
	{\bf Sample} & {\centering \bf Description} \\
	\hline\hline
	$\rm S_{SF}$ & Star-forming sample consisting of all central galaxies at $z=0$
	with $M_*\geq 10^{8.5} \msun$ and $\ssfr \geq 10^{-2} \gyri$, and with $10\%$ outliers eliminated. \\ \hline
	$\rm S'_{SF}$ & The same as $\rm S_{SF}$ but without eliminating the outliers.\\ \hline
	$\rm S_{SF,\,z=z_0}$ & The same as $\rm S_{SF}$ but selected at $z=z_0$. \\ \hline
	$\rm S_Q$ & Quenched sample consisting of all central galaxies at $z=0$
	with $M_*\geq 10^{8.5} \msun$ and $\ssfr < 10^{-2} \gyri$. \\  \hline
	$\rm S_{Q,\,z=z_0}$ & The same as $\rm S_{Q}$ but selected at $z=z_0$. \\ \hline
\end{tabularx}
\label{tab:def-samples}
\end{table*}
\end{center}

\subsection{The galaxy samples}
\label{ssec:sample}

In this paper, we focus on the formation of central galaxies
at $z = 0$. We thus select all central galaxies (the ones 
hosted by central subhalos) in TNG and EAGLE.
The grey shade in Fig.~\ref{fig:sample-separation} show the galaxy 
distribution in the $(\log\,\Mstar,\,\log\,\ssfr)$ plane, 
where $\ssfr$ is the smoothed sSFR (see Sec.~\ref{ssec:sim}). 
It is clear that there are two distinct populations: 
a star-forming main-sequence in which the sSFR is 
high and almost independent of $\Mstar$; a quenched population 
with low sSFR for which the star formation activity 
may not even be resolved by the simulations. 
Since the quenching of galaxies in star formation 
is expected to be regulated by feedback processes, 
the presence of the two distinct populations indicates 
that the physical processes operating in them are different. 
Motivated by this, we separate galaxies into two samples
as specified below.
\bit 
\item The star-forming sample $\rm S_{SF}$. This sample includes 
galaxies at $z=0$ with $\Mstar \geq 10^{8.5} \msun$ and the 
smoothed $\ssfr \geq 10^{-2} \gyri$, but with all galaxies 
that lie outside the 90\% contour of the distribution 
in the $(\log\,\Mstar,\,\log\,\ssfr)$ plane eliminated. 
\item The quenched sample ${\rm S_Q}$. This includes all galaxies at $z=0$ 
with $\Mstar \geq 10^{8.5}h^{-1} \msun$ and  $\ssfr < 10^{-2} \gyri$.
\eit 
In some of the following analysis, where a complete sample is needed,
we use a sample $\rm S'_{SF}$, which includes all $z=0$ central galaxies 
with $\Mstar \geq 10^{8.5} \msun$ and $\ssfr \geq 10^{-2} \gyri$. 
We also have analysis for which the properties of galaxies at 
a higher redshift $z_0$ are needed. In such cases, we apply the same 
separation criteria to the galaxies at the desired redshift, 
and construct samples $\rm S_{SF,\,z=z_0}$ and $\rm S_{Q,\,z=z_0}$
accordingly.
In \S\ref{ssec:reason-to-quench} and \S\ref{ssec:relations-in-quench} 
we have to further divide these samples according to stellar mass.
We will describe the sub-samplings when they are used. We summarize the samples
used in this paper in Table~\ref{tab:def-samples}.
The two $z=0$ samples defined above are shown by blue and red contours in 
Fig.~\ref{fig:sample-separation}. 

We checked our results by using a higher 
$\Mstar$ limit and a different sSFR threshold for the separation for the 
two populations, and by excluding post merger systems. We  
found that our conclusions are not sensitive to the criteria adopted.

\begin{figure*}\centering
	\includegraphics[width=15cm]{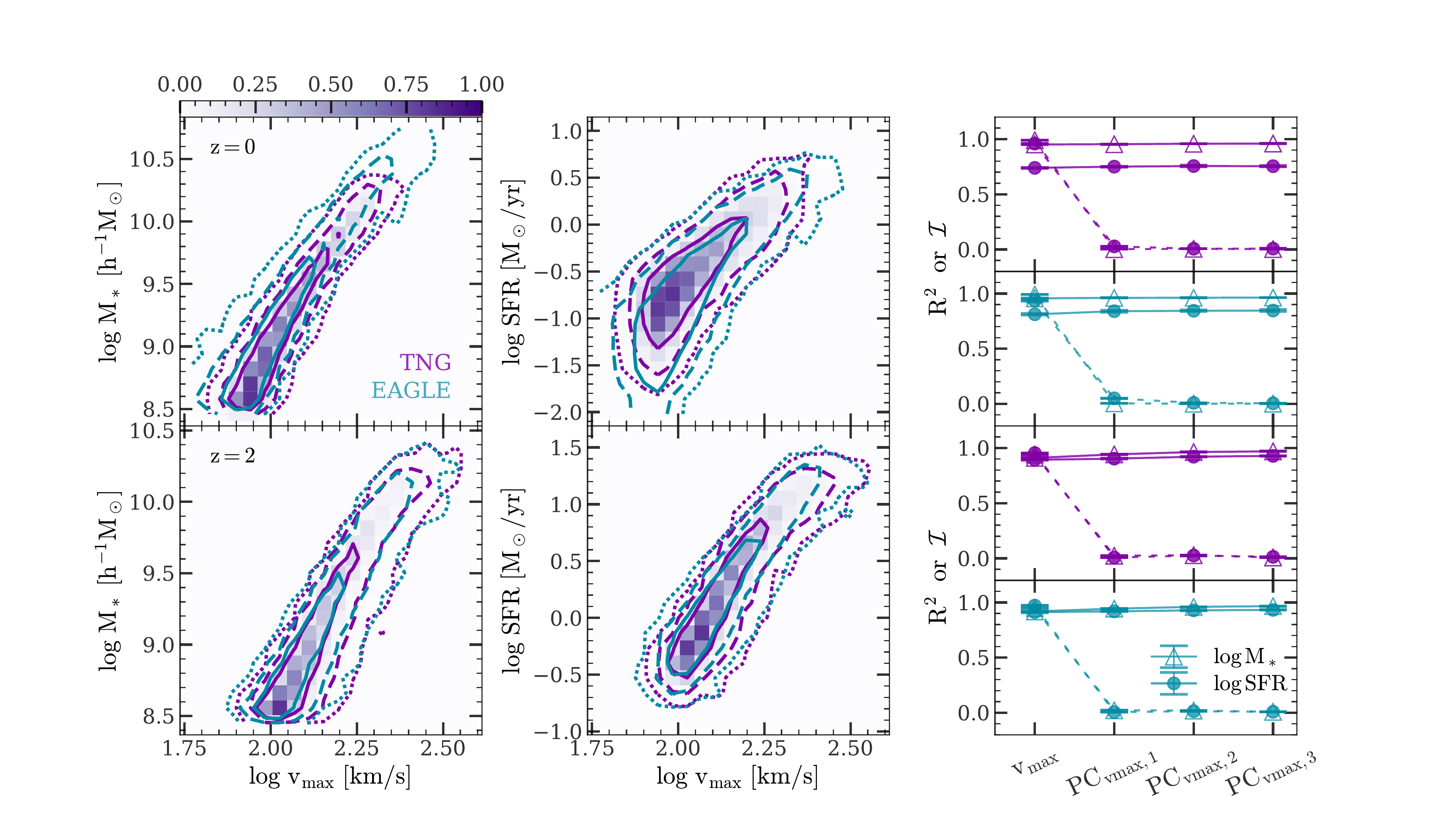}
	\caption{
		The halo-galaxy relations of the star-forming galaxies at two redshifts ({\bf upper} row, sample $\rm S_{SF}$;  {\bf lower} row, sample $\rm S_{SF, z=2}$). 
		In each row, the {\bf left} and {\bf middle} panel show the correlation
                of $\Mstar$ and the smoothed SFR with $\vmax$. {\rm Solid}, {\rm dashed} and {\rm dotted} contours enclose 1, 2 and 3$-\sigma$ regions, respectively.
		{\color{Purple}\bf Purple} shade represents the normalized distribution
                for TNG.
                The GBDT regression results are shown in
                the {\bf right} panels, where
                four halo properties are used to predict $\log\,\Mstar$ ({\bf triangulars}) 
                and $\log\,{\rm SFR}$ ({\bf circles}). 
		{\bf Solid} lines are cumulative $R^2$, and {\bf dashed} lines are the importance $\mathcal{I}$ of predictors in the regressor that uses the halo properties 
		labelled along the x-axis (see Appendix~\ref{app:GBDT} for the definitions of $R^2$ and $\mathcal{I}$).
		The error bars are computed by bootstrap resampling.
        Results from TNG and EAGLE are plotted in {\color{Purple} purple} and {\color{TealBlue}\bf green}, respectively.
	}
	\label{fig:sf-relations}
\end{figure*}

\begin{figure*}\centering
	\includegraphics[width=17cm]{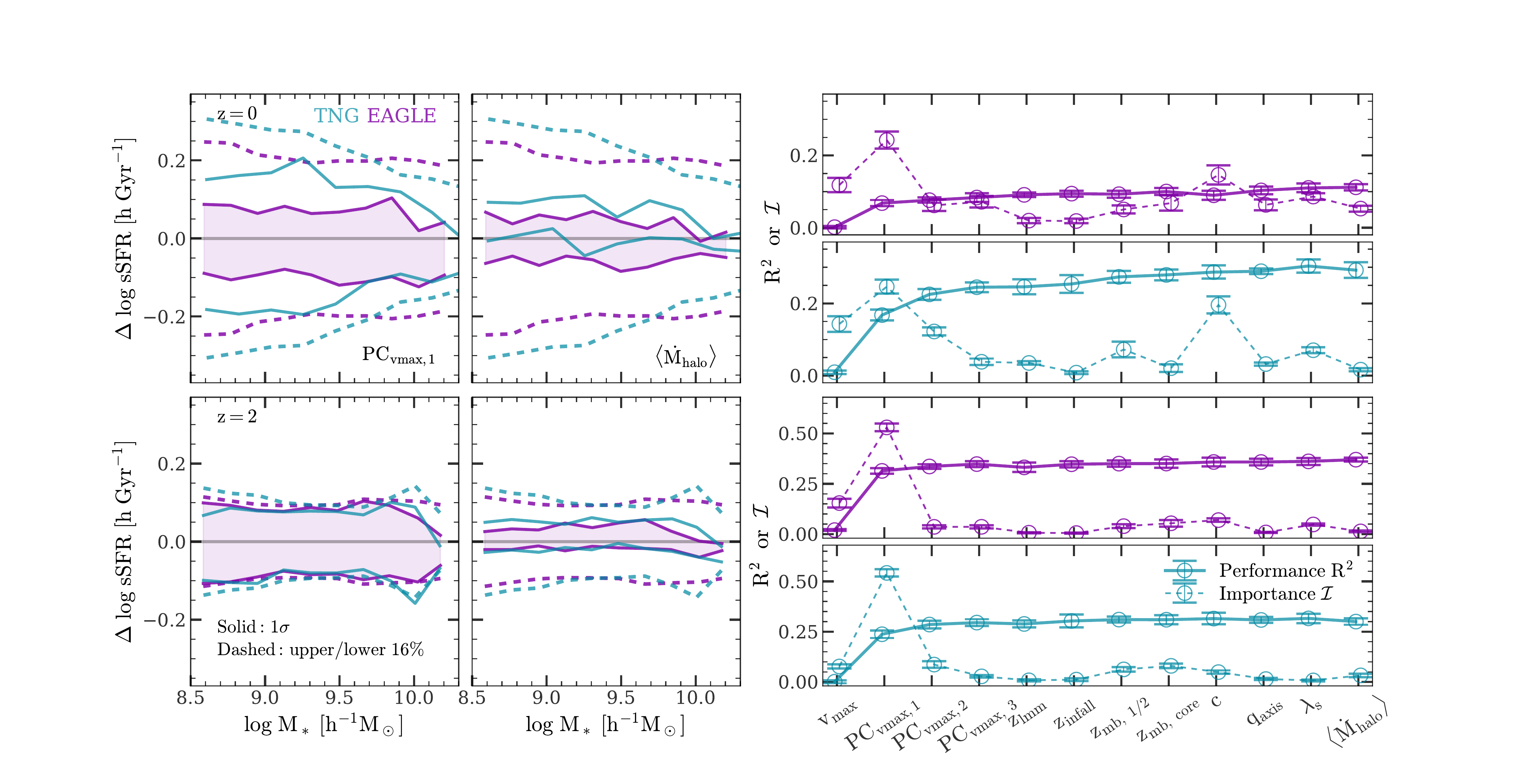}
	\caption{The two columns on the left display
	  the sSFR residual $\ssfrRes$ of star-forming galaxies as
          a function of $\log,\Mstar$ for sample $\rm S_{SF}$ 
          ({\bf upper} row) and sample $\rm S_{SF,z=2}$
          ({\bf lower} row). Results for TNG and EAGLE are plotted
          in {\color{Purple} purple} and {\color{TealBlue}\bf green}, respectively.
          In the first column, the {\bf solid} lines
          represent the mean value of $\ssfrRes$ for subsamples 
          of galaxies whose $\halovmaxpc[1]$ are among the
          highest $16\%$ and the lowest $16\%$ of the full sample, respectively.
          The {\bf solid} lines in the middle column show
          the results for subsamples selected by the halo accretion
          rate $\haloAccrRate$ instead of $\halovmaxpc[1]$.
          In both columns, the standard deviation of the full
          sample is plotted as  {\bf dashed} lines. 
          The {\bf right} panels show the results of the regression
          of $\ssfrRes$ on halo properties
          ({\bf solid}, cumulative $R^2$; {\bf dashed}, importance
          $\mathcal{I}$ of predictors in the regressor using
          halo properties indicated along the x-axis). 
          See Appendix~\ref{app:GBDT} for the definitions of $R^2$ and $\mathcal{I}$.
          Errors are computed by bootstrap resampling.
	}
	\label{fig:sf-scatter-regression}
\end{figure*}

\begin{figure*}\centering
	\includegraphics[width=17cm]{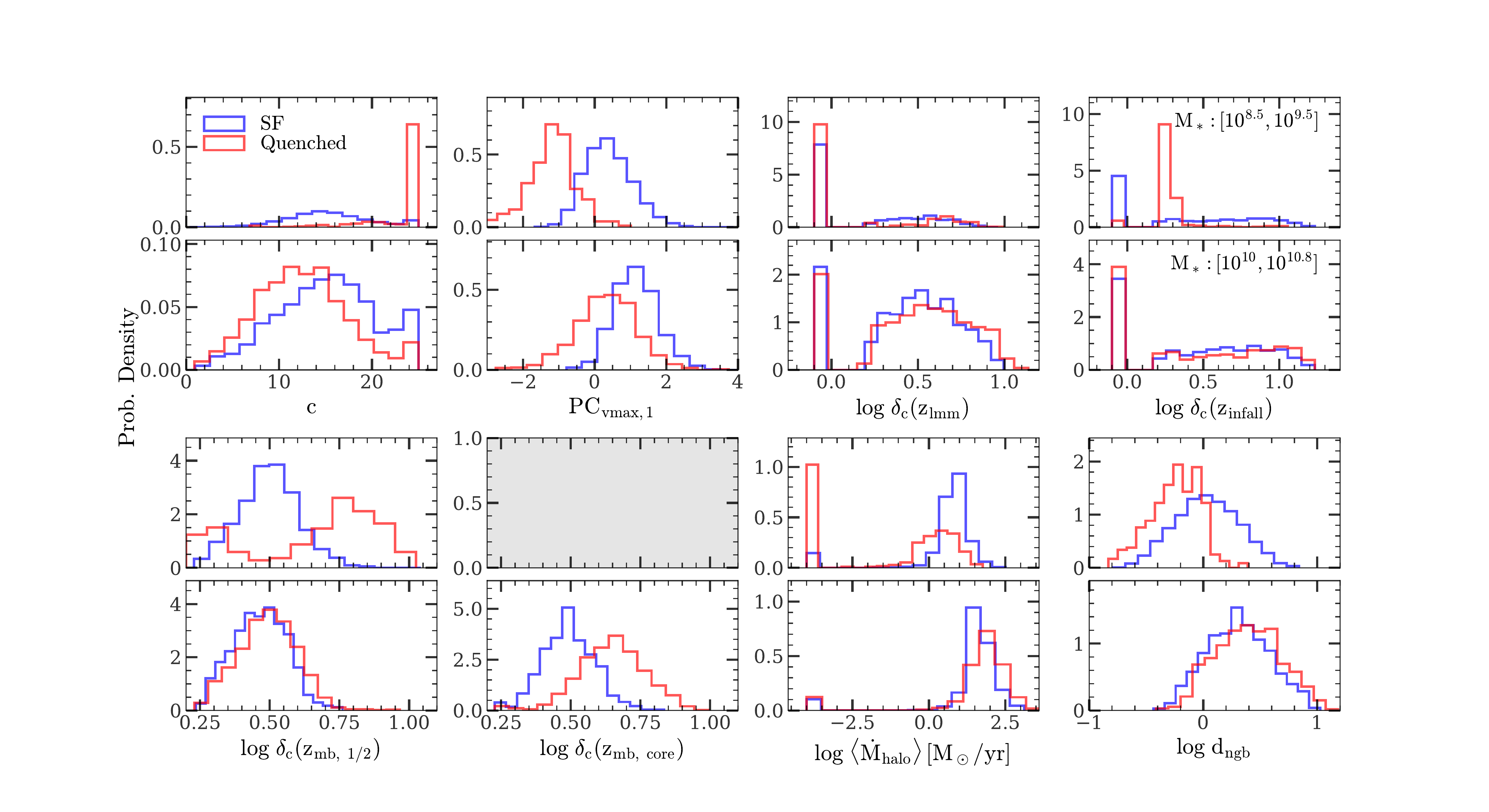}
	\caption{
	  The distributions of halo properties for TNG galaxies
          with ${\rm sSFR} \geq 10^{-2} {\rm Gyr}^{-1}h$
          ({\color{Blue}\bf blue}) and $<10^{-2} {\rm Gyr}^{-1}h$
          ({\color{Red}\bf red}). Each of the eight vertical pairs of 
          panels shows the distribution of one halo property. 
          In each pair, the {\bf upper} panel is for low-mass galaxies 
          and the {\bf lower} is for high-mass galaxies, as
          indicated in the upper right pair of panels (in units of $\msun$).
          Low-mass galaxies do not have $z_{\rm mb,\,core}$ measurements,
          and so no result is shown for this property.
	}
	\label{fig:reason-to-quench}
\end{figure*}

\begin{figure*}\centering
	\includegraphics[width=18cm]{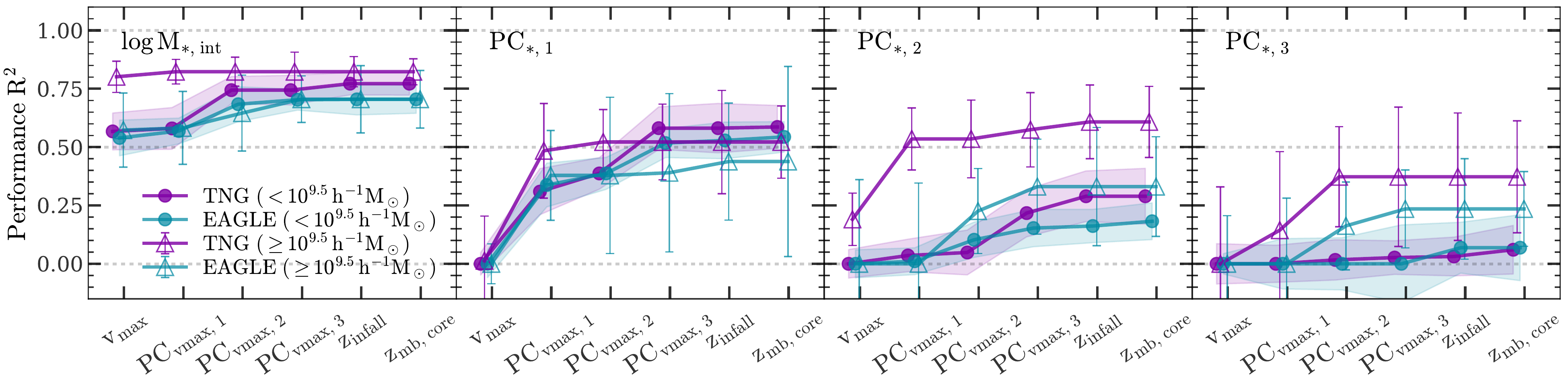}
	\caption{Cumulative $R^2$ of regressions of different stellar 
		properties on halo properties for the quenched galaxies (sample $\rm S_Q$). 
		In each panel, one galaxy property (as indicated at the upper left corner) is 
		regressed on six halo properties for both TNG ({\color{Purple}\bf purple}) 
		and EAGLE ({\color{TealBlue}\bf green}) and for both low-mass ({\bf circles}) 
		and high-mass ({\bf triangles}) galaxies. The stellar mass ranges ($M_*$) 
		are indicated in the first panel. 
        See Appendix~\ref{app:GBDT} for the definitions of $R^2$ and $\mathcal{I}$.
		Error bars and shaded regions are the
		errors estimated from bootstrap resampling.
	}
	\label{fig:quenched-relations}
\end{figure*}

\section{Relation between galaxy and halo properties in simulations}
\label{sec:galaxy-halo-conn}

Because of the bimodal distribution of galaxies as seen in 
\S\ref{ssec:sample}, we discuss the galaxy-halo relations 
separately for the star-forming and quenched populations. 
In this section, we first discuss the relation for the star-forming 
main sequence. We then examine how galaxies get quenched by looking 
at their halo properties. Finally, we present the galaxy-halo 
relation for the quenched population.

\subsection{Galaxy-halo relation for the main-sequence sample}
\label{ssec:relations-in-main-seq}

To quantify the correlation strength between halo and galaxy quantities, 
we use the model ensemble method GBDT to build a regressor, 
$y=f(\bfrm[x])$, which maps the set of halo quantities $\bfrm[x]$ to a 
galaxy quantity $y$. Using many predictor variables available, 
we can build a series of regressors with an increasing number of 
predictors. As the number of predictors increases, the overall 
performance, $R^2$, also increases. At each step, the amount of 
increase in $R^2$ caused by including a new variable $x \in \bfrm[x]$ 
can be used to judge whether this variable has any contribution to the target. 
When all of the predictors are included, the importance value, 
$\mathcal{I}$, output from the final regressor, can be used to judge 
the relative contributions from individual predictors.
The details of GBDT, {\bf $R^2$ and $\mathcal{I}$} can be found in Appendix \ref{app:GBDT}.

Fig.~\ref{fig:sf-relations} shows the galaxy-halo relations for the 
star-forming samples, $\rm S_{SF}$ and $\rm S_{SF,z=2}$,
in both TNG and EAGLE. Here the smoothed SFR and the first three PCs 
of the $\vmax$ history are used (see \S\ref{ssec:sim} and Appendix~\ref{app:PCA}). 
The results can be summarized as follows.
(i) For both low-z and high-z (the left and middle panels), 
both $\Mstar$ and the SFR are tightly correlated with $\vmax$. 
As shown in the right panel, even only $\vmax$ is used as 
the sole predictor, the value of $R^2$ is still quite large 
($\geq 0.9$ for $\Mstar$, $\geq 0.7$ for SFR at $z=0$ and 
$\geq 0.9$ for SFR at z=2). 
(ii) At z = 0, the relation between SFR and $\vmax$ has larger 
scatter than that between $\Mstar$ and $\vmax$. The smaller 
$R^2$ for SFR shown in the right panel also confirms this. 
This indicates that the factors regulating the star-formation 
activity becomes more diverse as galaxies evolve from high-$z$ 
to low-$z$. 
(iii) For both $\Mstar$ and SFR, and for both redshifts, adding more 
halo quantities into the predictor set does not significantly 
improve the regression performance $R^2$.
In all cases, $R^2$ is significantly larger than $50\%$ 
when only $\vmax$ is used. This indicates that the evolution 
of both $\Mstar$ and SFR is dominated by $\vmax$. The large 
contribution ($\mathcal{I}$) from $\vmax$ also validates this argument.

The tight $\Mstar$ - $\vmax$ and SFR - $\vmax$ relations indicate 
that the star-forming main-sequence is a well defined population
that is largely determined by the halo potential well 
represented by $\vmax$. Other halo properties, such as the MAH, 
are only secondary factors that produce relatively small variance 
in the sequence. To see which halo quantities are most responsible 
for the variance, we first define the residual value $\ssfrRes$ for 
the smoothed sSFR as follows: 
\bit
	\item  We build a GBDT regressor that maps $\log\, \Mstar$ to $\log\, \ssfr$. The predicted value of such a regressor is denoted as $\log\, \hat{\ssfr}(\log\,\Mstar)$, which can be viewed as the mean value of $\log\, \ssfr$ at a given stellar mass.
	\item We subtract the $\log\, \ssfr$ of each galaxy by the mean value at the corresponding stellar mass to get the residual,  $\ssfrRes=\log\,\ssfr-\log\,\hat{\ssfr}(\log\,\Mstar)$.
\eit
We relate the residual defined this way to halo quantities, as described below.

To see the effect of any halo property, $x$, on the main-sequence residual, 
we form two subsamples for galaxies of a given stellar mass. 
The first one consists of the 16\% with the highest $x$, while 
the second consists of the 16\% with the lowest $x$. 
If $x$ does have an effect on the variance of the main-sequence, 
these two subsamples should have different mean $\ssfrRes$. 
We do this for both the $\rm S_{SF}$ and $\rm S_{SF,z=2}$ samples using 
$x=\halovmaxpc[1]$, the first PC of the $\vmax$ history and 
$x=\haloAccrRate$, the halo accretion rate.  
The mean $\ssfrRes$ for the two subsamples at given stellar mass 
are shown in Fig.~\ref{fig:sf-scatter-regression} 
in comparison with the standard deviation of $\log {\rm sSFR}$ 
at the same stellar mass. Clearly, the means of $\ssfrRes$ in the two 
subsamples are different, and the effect of $\halovmaxpc[1]$ 
is significant in both TNG and EAGLE. Compared to the total 
main-sequence scatter, the effect appears relatively small 
at $z=0$ and becomes larger at higher $z$. 
Thus, using $\halovmaxpc[1]$ alone can only explain a small 
portion of the residual at $z=0$, and a larger portion at $z=2$. 
The $R^2$ values using only $\halovmaxpc[1]$ shown 
in the right panels are significantly less than $50\%$, 
confirming that the prediction power of $\halovmaxpc[1]$ is limited.
The results also show 
that the effect of $\haloAccrRate$ is smaller than that of 
$\halovmaxpc[1]$ at both $z=0$ and $z=2$ in both TNG and EAGLE. 
This indicates that the halo accretion rate is not as 
relevant as $\halovmaxpcs$ in affecting the star formation rate,
and is not a powerful proxy to separate galaxies according to
the sSFR galaxies. This is consistent with 
\cite{ODonnellC:BehrooziP:2020:Halo-accretion-rate-vs-SFR} 
who used the SDSS and an empirical model to demonstrate 
that the halo accretion rate does not significantly correlate with 
the current SFR, although some simulation-based investigations 
reached the opposite conclusion
\citep[e.g.,][]{WetzelAndrewR:2015:halo-accr-with-SFR}.

Again, because of the large number of halo quantities and 
potentially complex patterns in the feature space, 
we use GBDTs to relate the main-sequence residual, $\ssfrRes$,
to halo quantities. The cumulative $R^2$ from each of the 
regressors, and the contribution from each halo property in the 
final regressor using all halo properties, are shown in the right 
panel of Fig.~\ref{fig:sf-scatter-regression}. At $z=0$, 
the explained variance, i.e., $R^2$, is only about $10\%$ 
in TNG and about $30\%$ in EAGLE, even when a large set of halo 
properties are used. At $z=2$, $R^2$ for both TNG and EAGLE 
are still far less than $50\%$. These poor performances 
in terms of $R^2$ indicate that there is no dominant set 
of halo properties that can fully explain the variance in the 
main-sequence. Thus, once the main trend of SFR with respect
to the halo mass or to $\vmax$ is already taken into account, 
an empirical model should avoid using these halo properties
to assign the SFR of a galaxy based on deterministic ranking
or to direct predict the main-sequence residual. 
Part of the main-sequence residual has to be modeled 
as a random component with correct statistical properties. 
We will discuss how to build such a model in \S\ref{ssec:model}.

\subsection{Halo quantities that drive quenching}
\label{ssec:reason-to-quench}

Before moving to the quenched population, let us first 
examine why a galaxy is quenched. To be specific, we want to 
see which halo quantities can be used to predict whether a galaxy 
is quenched or not, and whether the prediction is deterministic 
or stochastic. This is crucial to empirical modeling. For example, 
in order for a halo-based model to predict the correct 
bimodal distribution for galaxies, we need a careful
model design so that the halo properties can really be used to 
distinguish between the star-forming and quenched populations.
Since low-mass galaxies and massive ones may be quenched through 
different processes (for example, supernova feedback may be more 
efficient in a low-mass galaxy, while AGN feedback may be stronger 
in a massive galaxy that can host a more powerful central SMBH), 
it is necessary to answer these questions separately for galaxies 
with different masses. We therefore define four sub-samples
according to both $\Mstar$ and the smoothed $\ssfr$, among all 
of the $z=0$ TNG galaxies. First, we separate these galaxies into 
two sub-samples with $ 10^{8.5} \leq \Mstar/(\msun) \leq 10^{9.5}$ 
and $ 10^{10} \leq \Mstar/(\msun) \leq 10^{10.8} $, respectively. 
We then split each of the two sub-samples into two subsets according 
to $\ssfr \geq 10^{-2} \gyri$ and  $\ssfr < 10^{-2} \gyri$, 
respectively. These four sub-samples are referred as the 
low-mass active sample, low-mass passive sample, 
high-mass active sample and high-mass passive sample, respectively. 
The choice of the stellar mass intervals is a compromise 
between minimizing the effect of $\Mstar$ and preventing each 
sub-sample from being too small.
 
For each of these four samples, we plot the the distributions of 
different halo quantities in Fig.~\ref{fig:reason-to-quench}. 
Here, $\log \delta_c(z_{\rm lmm})$ for a halo without any major 
merger is set to be a small negative value, and the same applies to 
$z_{\rm infall}$. The value of $\haloAccrRate$  
is set to be $10^{-4} \Msun/{\rm yr}$ when the measured value 
is small or negative. Galaxies in the low-mass sub-samples do not 
have the $z_{\rm mb,\,core}$ measurement, and so they do not appear 
in the panel of $z_{\rm mb,\,core}$.

For low-mass galaxies, the active and passive populations have totally 
different distributions in $z_{\rm infall}$. The active population 
has a flat $z_{\rm infall}$ distribution, while the passive one has a 
sharply peaked distribution. This difference strongly suggests that 
passive low-mass galaxies have undergone a very recent infall-ejection  
process, while high-mass galaxies do not. The distributions of 
other halo properties confirm this. For example, 
passive galaxies on average have smaller $d_{\rm ngb}$, 
consistent with the fact that the distance of such a galaxy 
to a massive halo must be small for the infall event to occur. 
Due to interactions in the infall-ejection process, the 
MAH of the halo can change significantly, 
which may change the distributions in  
$\halovmaxpc[1]$, $z_{\rm mb,\,1/2}$, and $\haloAccrRate$. 
The halo density profile may also change in this process, 
which explains why the distribution of $c$ for passive galaxies 
is also distinct from that of the star-forming population.

Although the distributions of halo properties for the 
star-forming and passive populations are significantly different, 
it is still challenging to design an ideal classifier to tell whether 
a low-mass galaxy is quenched or not using halo properties alone. 
The problem lies in the sample imbalance: the fraction of 
the passive population among all low-mass galaxies is less than 
$3\%$ in TNG, and less than $7\%$ in EAGLE. No matter how the 
classification boundary is drawn, there is always a large 
contamination in the population classified as the quenched population 
by star-forming galaxies.

The situation for high-mass galaxies is more complicated. 
Among all of the halo properties shown in Fig.~\ref{fig:reason-to-quench}, 
the only three which show large differences between star-forming and 
passive populations are $z_{\rm mb,\,core}$, $\halovmaxpc[1]$ and $c$. 
Because halos with mass larger than $M_{\rm h,\, core}$ 
may likely contain bright AGNs to quench star formation 
and be more dominated by hot model accretion, an earlier 
formation of a large progenitor, i.e. a higher $z_{\rm mb\, core}$,  
may be indicative of a higher probability for the galaxy 
to quench. Indeed, we can see this in the distribution of 
$z_{\rm mb,\, core}$, and, implicitly, in the 
distributions of $\halovmaxpc[1]$.
Compared with the quenched galaxies, the host halos of 
star-forming galaxies are more concentrated, because halos 
with smaller $z_{\rm mb,\,core}$ are less massive 
\citep{LiYun_MoHoujun_2008_HaloFormationTimesDef} and 
therefore more concentrated.
However, the distribution in $z_{\rm mb\,core}$, 
$\halovmaxpc[1]$ or $c$ has significant overlap between the 
star-forming and quenched populations. 
Thus, even though the star-forming and passive samples are more 
balanced for massive galaxies than for low-mass ones, 
it is still difficult to distinguish the two populations
for individual galaxies on the basis of the properties of their 
host halos. We also try to distinguish the star-forming 
and quenched populations by building GBDTs and using the 
combination of multiple halo quantities as features, 
including $M_{\rm mb,\,core}$, $\halovmaxpc[1]$ and $c$, 
with their effects shown in Fig.~\ref{fig:reason-to-quench},
and $\Mhalo$ and $\vmax$. Because of the degeneracy    
between halo properties, we find that including all these 
features makes no obvious improvement over using only 
$M_{\rm mb,\,core}$. These indicate again that a halo-based 
empirical model may not be able to predict galaxy quenching 
per individual halo. What we can do is to build a model 
capable of correctly predicting the statistical properties 
of galaxies for a large ensemble of halos.

\subsection{The relations in the quenched sample}
\label{ssec:relations-in-quench}

Once a galaxy is quenched, its SFR is lower and may even become too 
low to be resolved by the simulations. However, 
as we see from Fig.~\ref{fig:histories}, a quenched galaxy may 
have a high SFR in the past when it was still in the main sequence. 
The tight main-sequence relation seen in  
\S\ref{ssec:relations-in-main-seq}, therefore,  
indicates that the SFH of a galaxy 
may be related to the MAH of its host halo. 
Motivated by this, we use the integrated stellar mass, 
$\bfrm[M]_{\rm *,\,int}$, to represent the galaxy SFH. 
Compared to the current stellar mass, $\MstarInt$ at any 
given redshift can be viewed as the stellar mass that has formed 
throughout the history before the redshift (see \S\ref{ssec:sim} 
for definition). We have tested that our conclusion does not 
depend on this choice, because almost all galaxies in 
samples $\rm S_{SF}$ and $\rm S_Q$ have 
$\Mstar \leq \MstarInt \leq 2\Mstar$.

So defined, the $\MstarInt$ history is a direct quantity that 
``remember'' the history of star formation of a galaxy. Therefore, 
the $\MstarInt$ history should be connected to the halo MAH. 
Using the same PCA as used for the $\vmax$ history, we reduce the 
dimensions of the $\MstarInt$ histories by representing 
them with several PCs, denoted as 
$\bfrm[PC]_{*} = ({\rm PC}_{*,1},\,{\rm PC}_{*,2},\,...)$ 
(see \S\ref{ssec:sim} and Appendix~\ref{app:PCA}).

We relate $\MstarInt$ and $\rm PC_{*}$ of the quenched sample, 
$\rm S_Q$, to the following set of halo properties: $\vmax$, 
the PCs of the $\vmax$ history, $z_{\rm infall}$ and 
$z_{\rm mb,\,core}$, applying the GBDTs separately for 
low-mass ($M_* < 10^{9.5}\msun$) and high-mass 
($M_* \geq 10^{9.5} \msun$) sub-samples.
The inclusion of $z_{\rm infall}$ and $z_{\rm mb,\,core}$,
is motivated by the results presented in \S\ref{ssec:reason-to-quench},
where it is shown that these two quantities are responsible for the 
quenching of low-mass and high-mass galaxies, respectively. 
The results are shown in Fig.~\ref{fig:quenched-relations}. 
The prediction of $\MstarInt$ has a 
$R^2$ larger than $50\%$, even if
we use only $\vmax$ as the predictor. 
This indicates that $\MstarInt$ can be well reproduced with halo 
properties, and is dominated by $\vmax$. 
The SFH PCs are harder to predict. By using all of the 6 halo quantities, 
the $R^2$ for each of the three SFH PCs is still far less than $100\%$, 
indicating that a large portion of factors affecting the SFH
are still missing in the model and that the details
of how a galaxy forms may be influenced by many nuanced factors.
For the $\rm PC_{*,1}$, $R^2$ is significant and 
$\halovmaxpc[1]$ is the most important factor, 
as seen from the big increase of the cumulative $R^2$ it 
produces. 
The second and third PCs of 
the $\vmax$ history are also important for the $\rm PC_{*,1}$ of 
low-mass galaxies. As shown in \S\ref{ssec:reason-to-quench}, 
the quenching of low-mass galaxies is mainly due to the 
infall process in which their SFHs have large variances 
and more halo PCs are needed to capture them. 
For the $\rm PC_{*,2}$ and $\rm PC_{*,3}$ of high-mass galaxies, 
the TNG and EAGLE show large differences. The $R^2$ for EAGLE is 
much lower, and require more halo PCs to capture the variances. 
For low-mass galaxies, both TNG and EAGLE have low $R^2$
for the predictions of $\rm PC_{*,2}$ and $\rm PC_{*,3}$, 
indicating that high-order variations in the SFH are 
typically more difficult to model. In all the cases, $z_{\rm infall}$ 
and $z_{\rm mb,\,core}$ do not provide a significant contribution to 
$R^2$ when PCs of the $\vmax$ history are already used. 
This indicates that information carried by these two characteristic 
redshifts are already contained in the PCs of the $\vmax$ history. 
We will use these results to help build our empirical model, 
as described in \S\ref{ssec:model}.

\begin{figure*}\centering
	\includegraphics[width=18cm]{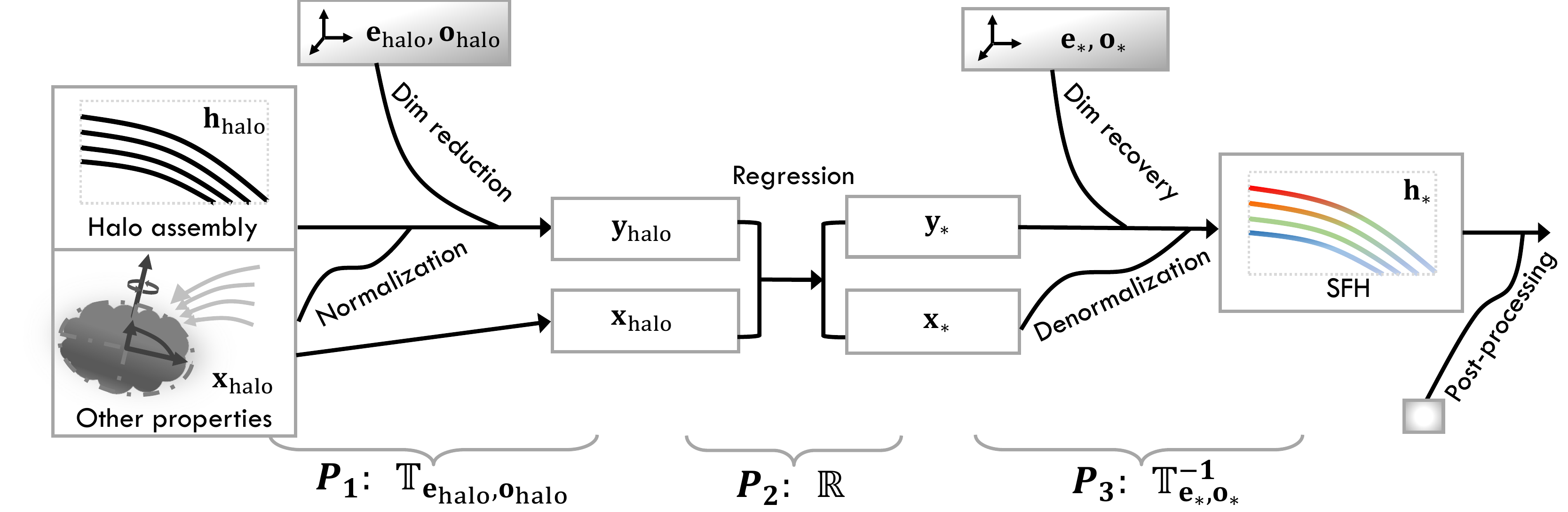}
	\caption{
	  The outline of the empirical model for the star formation of
          central galaxies in dark matter halos.
          The MAH, $\bfrm[h]_{\rm halo}$, and other properties of halos,
          $\bfrm[x]_{\rm halo}$, are transformed into the galaxy star formation
          history, $\bfrm[h]_*$, through three procedures, $P_1$, $P_2$ and $P_3$.
          A post-processing is performed in the end.
          See \S\ref{ssec:model} for the detailed description of the model.
	}
	\label{fig:sf-model}
\end{figure*}

\section{The empirical model of star formation in dark matter halos}
\label{sec:model}

The results presented in \S\ref{ssec:relations-in-main-seq} and 
\S\ref{ssec:relations-in-quench} show that the properties of the 
SFH of a central galaxy is well captured by the halo properties 
when the galaxy is in the star-forming main sequence, and that the 
$\MstarInt$ history can be well captured by halo properties even 
for quenched galaxies. Based on this tight galaxy-halo connection, 
we propose an empirical model to populate halos with central galaxies. 
Because of the differences between the star-forming and 
quenched populations, we model them separately.
In this section, we first discuss the design of the model and 
present a detailed description of all of the model ingredients. 
We then use five cases to test the model step by step. 

\subsection{The model}
\label{ssec:model}

The structure of the model is designed on the basis of 
the following considerations. 
(i) We favor a simple model with 
a small number of parameters to a complicated black-box model. 
A simpler model is easier to understand, and can provide 
more transparent insights into the relation between 
galaxies and dark matter halos. In addition, a simpler model is 
less prone to over-fitting problems. We thus choose the use of PCA to reduce 
the size of the parameter space for both halos and galaxies. 
(ii) The model should be expressive and flexible to absorb 
a variety of observation constraints and to provide a wide range 
of outcomes to compare with future observations. 
Because of this, we choose to build the model in a deeper way 
rather than directly mapping halo properties to galaxy properties. 
The model should include the full pipeline of the feature extraction, 
the regression, and the post-processing, each of which is simple 
enough while the joint of them is sufficient to capture the 
complicated patterns in the galaxy-halo connection. 
(iii) To optimize such a model, no standard approach is available.
Here we choose to break the model into several pieces and optimize 
them step-wisely. This optimization borrows the idea from the 
``greedy algorithm'' described in many textbooks of 
algorithm-design \citep[e.g., ][]{CormenThomasH:2009:IntroductionToAlgorithms, SedgewickR:2011:Algorithms}.

We outline the model in Fig.~\ref{fig:sf-model}. The overall purpose of 
the model is to predict the SFH, $\hstar$, and other properties, 
$\xstar$, of a given central galaxy, from the MAH of its host halo, 
$\hhalo$, and a set of other halo properties, $\xhalo$. 
Here $\xhalo$ and $\xstar$ are defined at a given redshift $z_0$, 
and $\hhalo$ and $\hstar$ are histories defined over a redshift 
range between $z_0$ and $z_1>z_0$. 
We will specify the definitions of these variables later. 
To achieve our goal, we break the model construction into three 
procedures, $P_1$, $P_2$ and $P_3$, which are described one-by-one 
in the following.

In the first procedure, $P_1$, we reduce the dimension of the halo 
MAH, $\bfrm[h]_{\rm halo}$. The purpose is to make the 
representation of a halo simpler so that the mapping from it to galaxy 
properties is easier to establish. $P_1$ consists of the following steps.
\bit 
	\item We choose $\hhalo=\bfrm[v]_{\rm max}$ as the halo MAH 
	variable and use only $\vmax$ for $\xhalo$, 
	$\bfrm[x]_{\rm halo}=(\vmax)$. 
	We have checked other halo properties, such as halo virial mass 
	and mass bound to subhalos and found that $\vmax$ is the best. 
	This is consistent with the test results of subhalo abundance matching in 
	\cite{ReddickRM:WechslterRH:2013:SHAM:VmaxBest}, 
	but here we extend it by including the MAH as a secondary halo property.
	We choose only PCs of the $\vmax$ history as the history variable, 
	because we have already seen that the $\vmax$ history is 
	tightly related both to the SFR for galaxies in the main sequence, 
	and to the history of $\MstarInt$ even for quenched galaxies.
	
	\item We normalize the halo MAH by $\tilde{\bfrm[h]}_{\rm halo}
	=\hhalo/h_{\rm halo,\,z=z_0}$, where $h_{\rm halo,\,z=z_0}$ is 
	the component of $\bfrm[h]_{\rm halo}$ that corresponds to $z=z_0$. 
	The purpose is to prevent the dimension reduction from 
	being too much concentrated in low redshift.
	
	\item We apply the PCA to $\tilde{\bfrm[h]}_{\rm halo}$, which gives 
	a set of eigenvectors, $\bfrm[e]_{\rm halo,\,i}\ (i=1,2,3,...)$, 
	and a mean offset, $\bfrm[o]_{\rm halo}$ (see Appendix~\ref{app:PCA}). 
	After a shift of $\bfrm[o]_{\rm halo}$ and a projection 
	with $\bfrm[e]_{\rm halo}$, we get a new vector $\bfrm[y]_{\rm halo}$, 
	which is the set of PCs we want to obtain.
	Our test shows that using the first two PCs is sufficient for modeling 
	the SFH, and that including more PCs does not lead to much gain in the model 
	performance.
\eit 

After procedure $P_1$, a halo can be described by a small set of 
variables $(\bfrm[y]_{\rm halo}, \bfrm[x]_{\rm halo})$, which is sufficiently 
simple. As shown in \S\ref{ssec:relations-in-main-seq} and 
\S\ref{ssec:relations-in-quench}, this set also gives a good prediction 
for the SFH. We denote the total transformation in procedure $P_1$ as $\mathbb{T}_{\bfrm[e]_{\rm halo},\bfrm[o]_{\rm halo}}$: 
\begin{equation}
	(\bfrm[y]_{\rm halo},\,\bfrm[x]_{\rm halo}) = \mathbb{T}_{\bfrm[e]_{\rm halo},\bfrm[o]_{\rm halo}}(\bfrm[h]_{\rm halo},\, \bfrm[x]_{\rm halo}) \,,
\end{equation}
where $\bfrm[x]_{\rm halo}$, which is not involved in the transformation, 
is included to simplify descriptions in the following. 
The halo properties $(\bfrm[y]_{\rm halo},\,\bfrm[x]_{\rm halo})$ are then fed into procedure $P_2$.

Before entering $P_2$, we must decide how to represent a galaxy. 
One of the quantities of interest is $\MstarInt$, and 
we denote the set of stellar properties as $\bfrm[x]_*=(\MstarInt)$
in this case. The SFH is a large vector, too complicated to model. 
It is therefore necessary to represent the SFH also by a set of PCs. 
For both star-forming and quenched galaxies, the SFH is well 
correlated with halo properties (see \S\ref{ssec:relations-in-main-seq} 
and \S\ref{ssec:relations-in-quench}),
so we define the SFH as $\bfrm[h]_*=\log\,\bfrm[M]_{\rm *,\,int}$.
The normalization is performed as 
$\tilde{\bfrm[h]}_*=\bfrm[h]_* - h_{\rm *,\,z=z_0}$. 
Note that we also tested using other galaxies properties to represent SFH, e.g., 
$\bfrm[SFR]$ for star-forming galaxies, but found little difference
in terms of the model performance.
Once $\tilde{\bfrm[h]}_*$ is obtained, we use the same method as we did 
for halos to reduce the dimension of SFH into a set of PCs, $\bfrm[y]_*$, 
given by the set of eigenvectors, $\bfrm[e]_{*,\,i}(i=1,2,3,...)$, 
and the mean offset, $\bfrm[o]_*$. The normalization and the projection 
into the new frame are jointly referred as the transformation 
$\mathbb{T}_{\bfrm[e]_*,\,\bfrm[o]_*}$, so that 
$(\bfrm[y]_*,\,\bfrm[x]_*)=\mathbb{T}_{\bfrm[e]_*,\,\bfrm[o]_*}(\bfrm[h]_*,\, \bfrm[x]_*)$.
The real modeling process is actually the inverse, namely
we first predict the PCs of the SFH and $\bfrm[x]_*$ according to 
halo properties, and then do the reverse transformation to obtain 
the SFH. This is what we do in procedures $P_2$ and $P_3$.

Procedure $P_2$ is simple: we build a regressor to predict stellar 
properties of a galaxy, $(\bfrm[y]_*,\,\bfrm[x]_*)$, according to 
halo properties. Denoting the regressor as $\mathbb{R}$, we have 
\begin{equation}
	(\bfrm[y]_*,\,\bfrm[x]_*) = \mathbb{R}(\bfrm[y]_{\rm halo},\,\bfrm[x]_{\rm halo})  \,.
\end{equation}
Procedure $P_3$ is just the reverse of 
$\mathbb{T}_{\bfrm[e]_*,\,\bfrm[o]_*}$:
\begin{equation}
	(\bfrm[h]_*,\, \bfrm[x]_*) = \mathbb{T}^{-1}_{\bfrm[e]_*,\,\bfrm[o]_*}(\bfrm[y]_*,\,\bfrm[x]_*)\,,
\end{equation}
which includes the dimension recovering and denormalization. 
The dimension recovering transforms $\bfrm[y]_*$ back to $\tilde{\bfrm[h]}_*$ by 
the inverse of the PCA using $\bfrm[e]_*$ and $\bfrm[o]_{*, i}$. 
The de-normalization transforms $\tilde{\bfrm[h]}_*$
to $\bfrm[h]_*$ using $\bfrm[x]_*$. Note that $\bfrm[x]_*$ is not 
changed in the transformation.

Putting all these procedures together, we have a mapping from halo properties, 
$(\bfrm[h]_{\rm halo},\,\bfrm[x]_{\rm halo})$, to galaxy properties, 
$(\bfrm[h]_*,\, \bfrm[x]_*)$:
\begin{equation}
	(\bfrm[h]_*,\, \bfrm[x]_*) = 
	\mathbb{T}^{-1}_{\bfrm[e]_*,\,\bfrm[o]_*}\,\mathbb{R}\,\mathbb{T}_{\bfrm[e]_{\rm 
	halo},\bfrm[o]_{\rm halo}}(\bfrm[h]_{\rm halo},\,\bfrm[x]_{\rm halo}) \,.
\end{equation}
Note that the model has some degrees of freedom to be fixed. 
The dimension reduction templates for halos, $\bfrm[e]_{\rm halo, i}$ and $\bfrm[o]_{\rm halo}$, 
are always known because we populate halos in 
dark-matter simulations. On the other hand, the regressor, $\mathbb{R}$, 
in $P_2$ needs to be modeled for real applications. The template for galaxies, 
$\bfrm[e]_*$ and $\bfrm[o]_*$, also needs to be modeled. 
The main advantage of our model is that, we may borrow some unknown 
parts from hydrodynamic simulations, so that the degrees of freedom 
of the model can be reduced dramatically. For example, 
although galaxy SFHs in different simulations may differ  
significantly, they may still be represented accurately 
by a small number of PCs with eigen-functions obtained 
from one set or a combination of multiple sets of 
simulations, thus reducing the dimension of each individual 
SFH from infinity (to describe a continuous history) to a small 
number. We will discuss the details in 
\S\ref{ssec:model-test} and show the results 
in \S\ref{ssec:model-result}.

Since we only take $\MstarInt$ at $z=z_0$ as the normalization for galaxy SFH, 
a small discrepancy in the prediction of $\MstarInt$ may give rise to 
a large difference in SFH at high redshift. To make the model more precise 
at high redshift, we break the halo MAH and the SFH of central 
galaxies at $z=0$ into two pieces: the first is between $z_0=0$ and 
$z_1=1.5$, and the second is above $z_0=1.5$. We run the model separately 
for these two redshift ranges, and join the modeled $\hstar$ with a proper 
smoothing at $z\sim 1.5$. This, of course, doubles the model complexity 
but gives a more accurate prediction of the SFH, which may be needed 
when high-$z$ data are available to constrain the model.

For any galaxy, once $\bfrm[M]_{\rm *,\,int}$ is known, we can differentiate it
with respect to cosmic time to obtain $\bfrm[SFR]$.

As shown in \S\ref{ssec:relations-in-main-seq}, 
the SFR of a galaxy cannot be totally determined by its halo 
properties even for galaxies in the main sequence.
The residual of the main sequence is hard to predict 
even with the use of a large set of halo properties. 
So far our model has not taken the scatter in the SFR into 
account. To make the model for the star-forming galaxies more realistic, 
we add a Gaussian random component with a zero mean and a 
covariance $\Sigma$ to the modeled $\log\,\bfrm[sSFR]$. 
The covariance is obtained by fitting that of the 
residual between the modeled $\log\,\bfrm[sSFR]$ and the 
simulated one. The logarithm of each diagonal element of 
$\Sigma$ is fitted by a sigmoid function $\sigma$ versus 
$\log(1+z)$, and each off-diagonal element is fitted by 
a linearly decayed correlation strength versus the number of 
snapshots between any two elements, $i-j$:
\begin{equation}
	\log\,\Sigma_{i,j}=\sigma[\log(1+z_i)]\sigma[\log(1+z_j)]{\rm Lin}(i-j),\, 
\end{equation}
where the four free parameters in the sigmoid function and the 
two free parameters in the linear function are all free parameters 
to be determined by the fit. We found that the correlation length 
is always quite small, with the correlation quickly decreasing 
to a negligible value. We also found that the sigmoid behavior of 
the variance does not depend strongly on halo mass.

All the processes after $P_3$ are collectively referred to as the 
post-processing.

\begin{figure*}\centering
	\includegraphics[width=16.5cm]{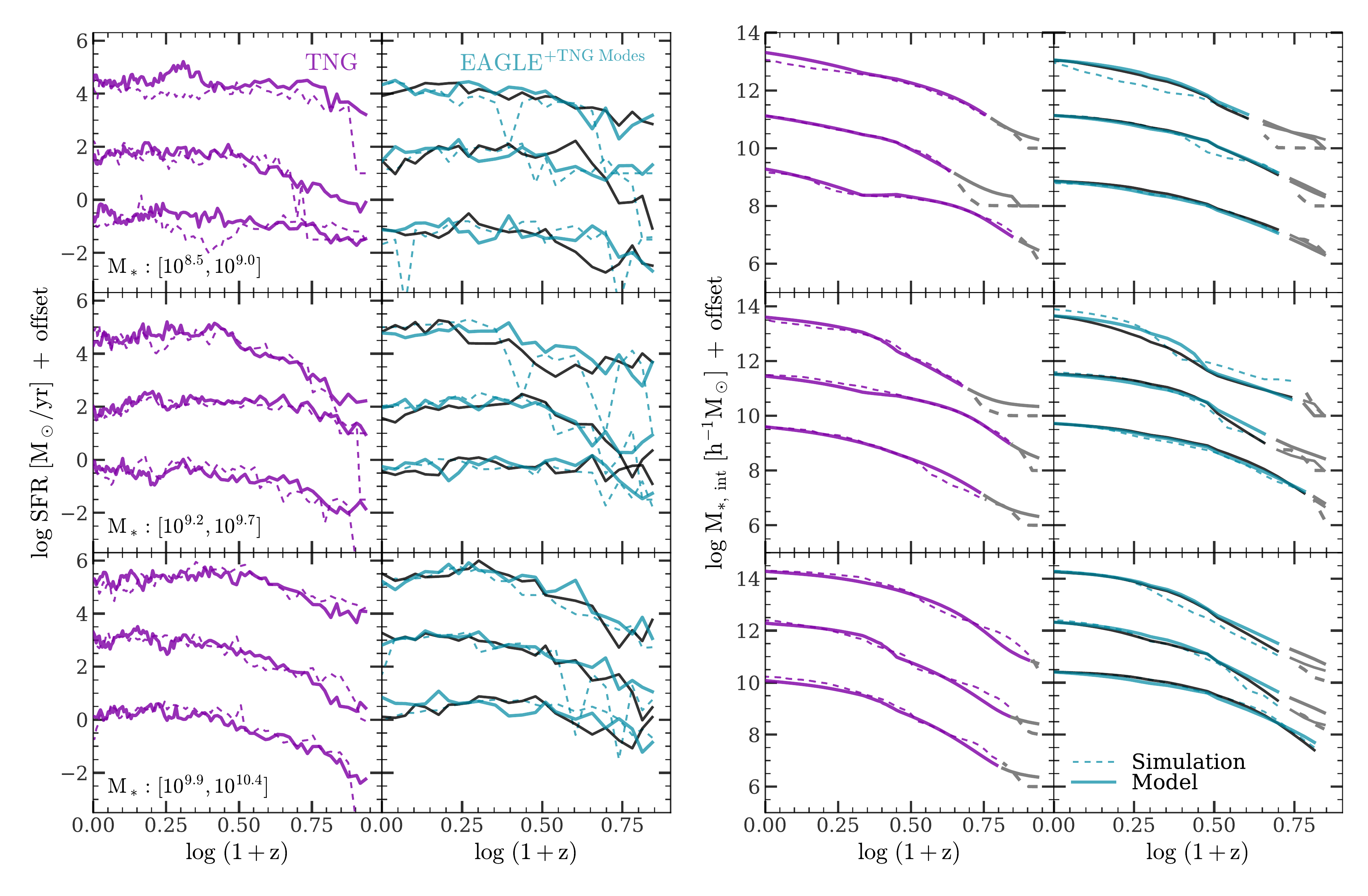}
	\caption{The modeled SFR and $\MstarInt$ histories of 
          star forming galaxies at $z=0$ (sample $\rm S_{SF}$) compared with
          the simulation results for three test cases,
          $\rm T_{TNG}$, $\rm T_{EAGLE^{+TNG-modes}}$ and $\rm T_{EAGLE}$,
          and for three different stellar mass ranges as indicated in
          the leftmost panels (in units of $\msun$).
          In each panel, {\bf solid} lines are from our model,
          while {\bf dashed} lines are from simulations.
          {\bf \color{Purple} Purple} lines are for the test case $\rm T_{TNG}$,
          {\bf \color{TealBlue} green} lines for
          $\rm T_{EAGLE^{+TNG-modes}}$, and {\bf black} lines, plotted
          together with the green lines, for $\rm T_{EAGLE}$.
          In each case, results are shown for three example galaxies,
          with arbitrary offsets vertically for clarity.
          {\bf \color{Gray} Gray} lines in the panels of $\MstarInt$ histories
          show the parts of the histories where the stellar mass 
          is below $10^7 \msun$ and which may not be well resolved in the 
          simulations. 
	}
	\label{fig:sf-model-history}
\end{figure*}

\begin{figure*}\centering
	\includegraphics[width=18cm]{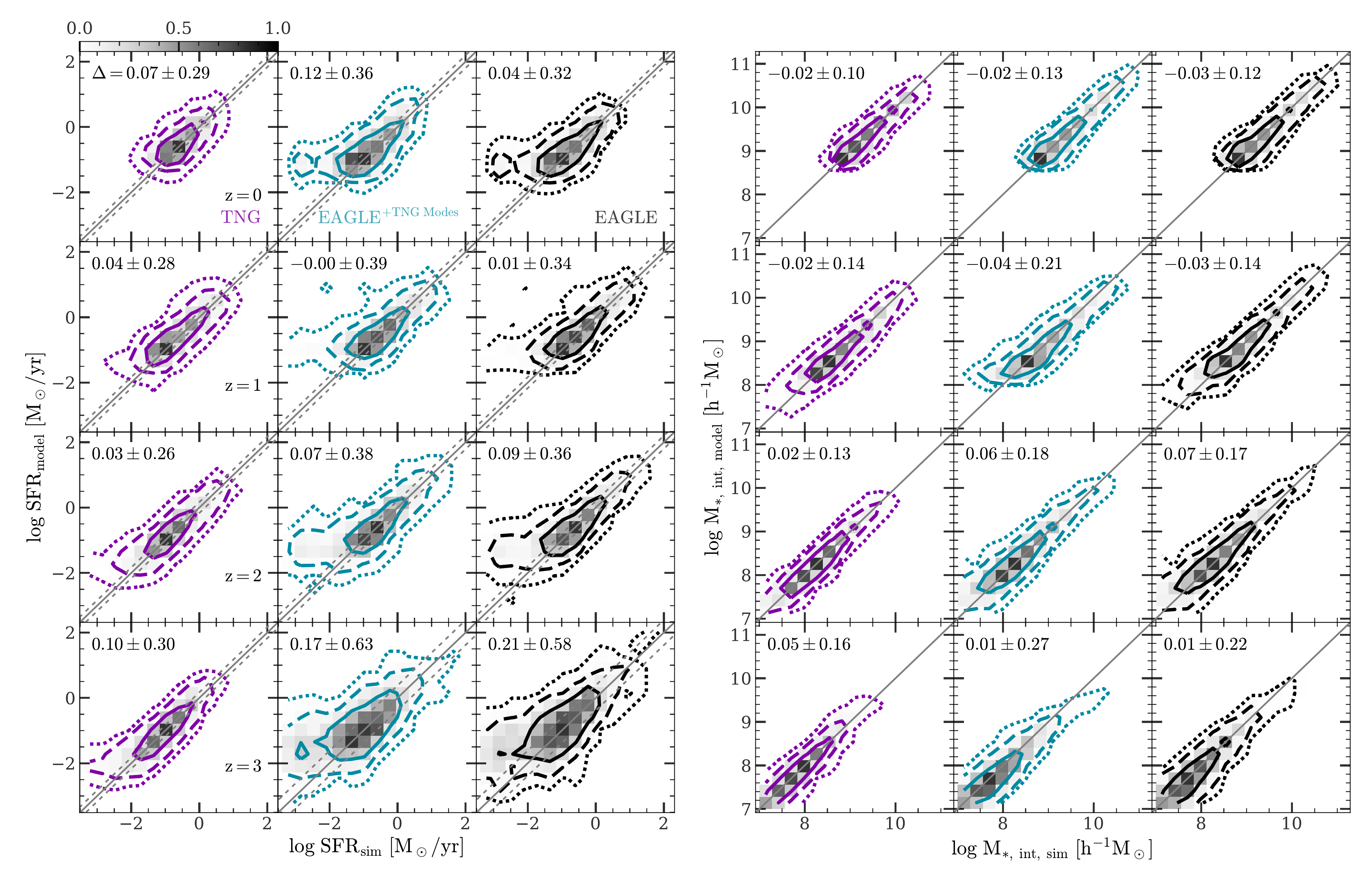}
	\caption{The modeled SFR (the left three columns) and $\MstarInt$ 
	(the right three columns) in the histories
          of $z=0$ star-forming galaxies (sample $\rm S_{SF}$)
          in comparison with the simulation results for three test cases,
          $\rm T_{TNG}$, $\rm T_{EAGLE^{+TNG-modes}}$ and $\rm T_{EAGLE}$,
          and for different redshifts as indicated in the left panels.
          The $M_*$-weighted average of residual and standard deviation
          ($\Delta$) are indicated in the upper left corner of each panel.
          The modeled SFR includes the randomly added noise whose standard
          deviation is shown by the dashed {\bf \color{Gray} gray} curves
          in each of the SFR panels. {\bf Solid}, {\bf dashed} and {\bf dotted}
          contours enclose 1, 2 and 3$-\sigma$ regions, respectively.
	}
	\label{fig:sf-model-pp-compare}
\end{figure*}

\begin{figure*}\centering
	\includegraphics[width=16.5cm]{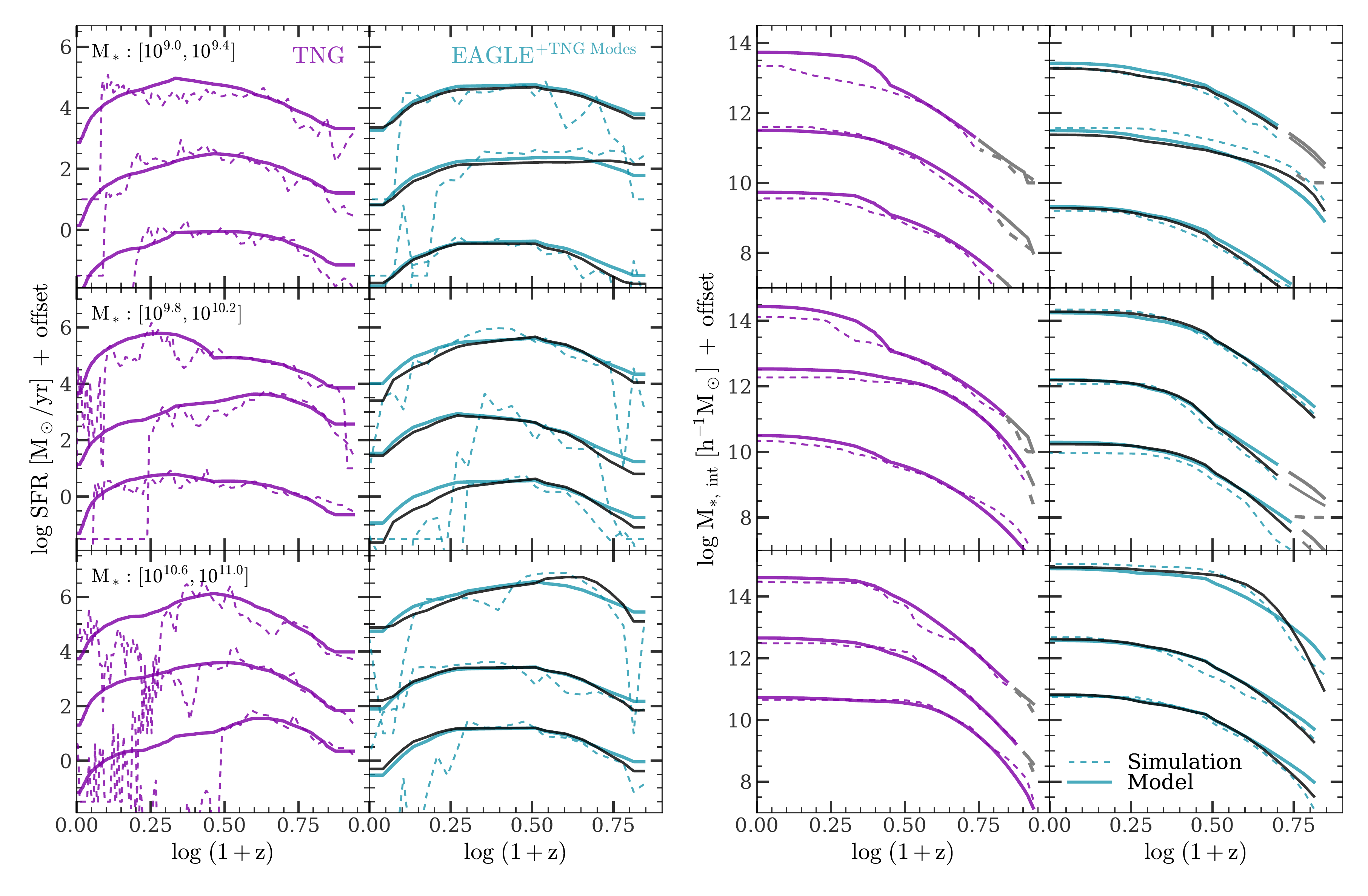}
	\caption{The same as Fig.~\ref{fig:sf-model-history}, but for the
          quenched galaxy sample $\rm S_Q$.
	}
	\label{fig:qch-model-history}
\end{figure*}

\begin{figure*}\centering
	\includegraphics[width=18cm]{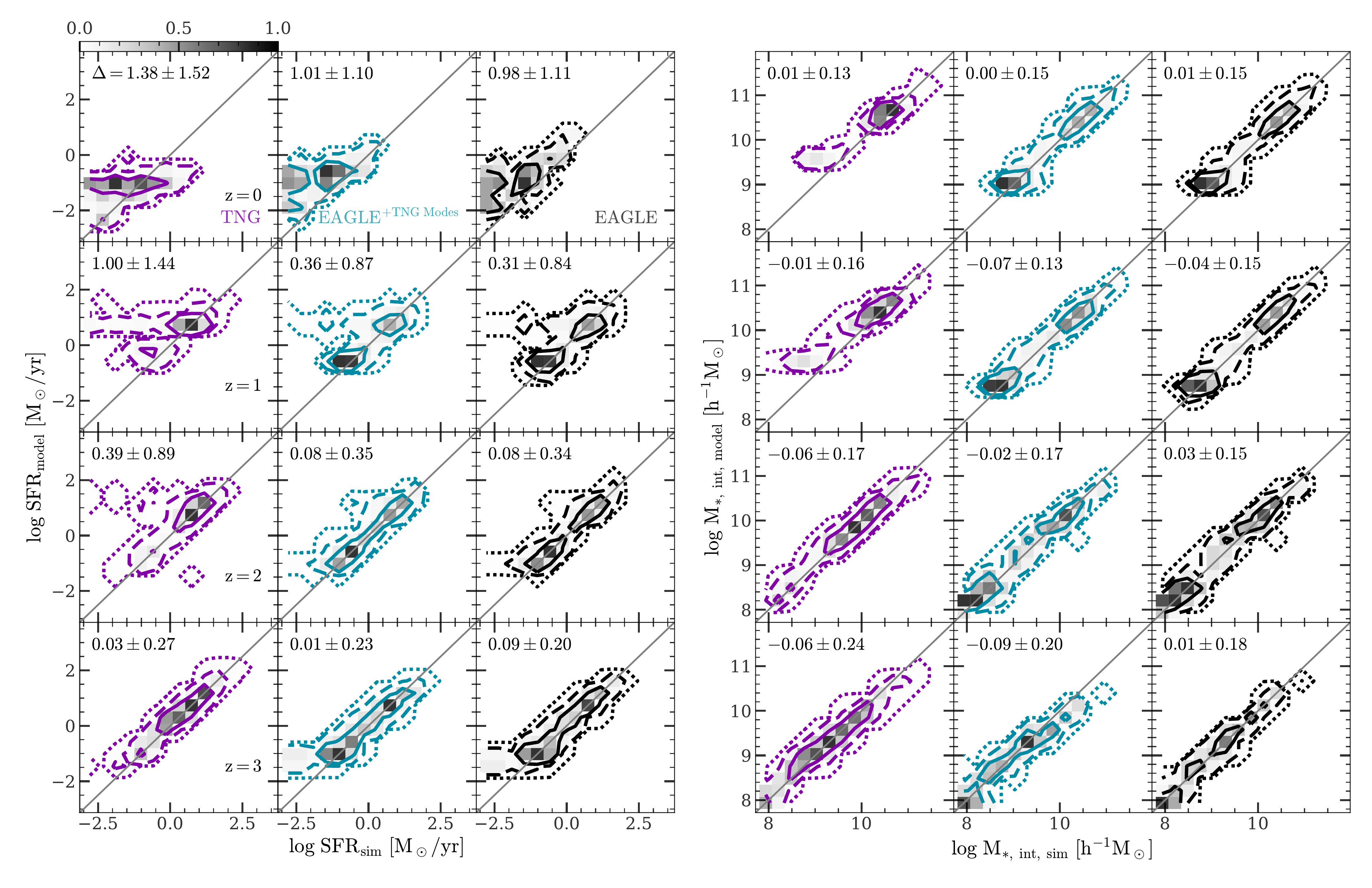}
	\caption{
	  The same as Fig.~\ref{fig:sf-model-pp-compare}, but for the
          quenched galaxy sample $\rm S_Q$.
	}
	\label{fig:qch-model-pp-compare}
\end{figure*}

\begin{center}
\begin{table*}
	\caption{ Five test cases for the empirical model used in this paper. 
		The exact definitions can be found in \S\ref{ssec:model-test}. This table lists the target galaxies we want to model and compare with the simulation, the simulation which the SFH template is taken from, and how to model the star-forming and quenched populations. }
	\begin{tabular} { c||c|c|c|c|c } 
		\hline
		Test Case & $\rm T_{TNG}$ &  $\rm T_{EAGLE^{+TNG-modes}}$ & $\rm T_{EAGLE}$
		& $\rm T_{join}$ & $\rm T'_{join}$ \\
		\hline\hline
		Target Galaxies & TNG & EAGLE & EAGLE & EAGLE & EAGLE \\
		\hline
		SFH Template & TNG & TNG & EAGLE & TNG & EAGLE \\
		\hline
		Treatment of Bimodal Populations & \multicolumn{3}{c|}{Separately} & \multicolumn{2}{c}{Jointly} \\
		\hline
	\end{tabular}
	\label{tab:def-test-cases}
\end{table*}
\end{center}

\subsection{Testing the model with simulations}
\label{ssec:model-test}

We now apply our model to simulations and test its performance by comparing 
the model prediction with the simulated SFH of galaxies. 
We define five test cases, denoted as $\rm T_{TNG}$, 
$\rm T_{EAGLE^{+TNG-modes}}$, $\rm T_{EAGLE}$, $\rm T_{join}$ 
and $\rm T'_{join}$, respectively. This design allows us to test our model 
both in ideal cases, where all of the model ingredients are known, 
and in more realistic cases, where some of the model ingredients 
need to be modeled. We summarize the test cases in Table~\ref{tab:def-test-cases}.

The test cases defined here use dark matter halos in full hydrodynamic runs. 
We checked our results by matching these subhalos with those in the corresponding dark-matter-only (DMO) runs, and rerunning our model on the DMO subhalos. 
We found no obvious changes in our results, 
although the uncertainty in the modeled stellar properties increases moderately. This may be 
expected, as the halo structure on the scale 
relevant to our modeling (e.g. where $\vmax$ is defined) may not be affected significantly by the baryonic effects. We should also emphasize that 
we use hydro simulations to guide our model design, 
rather than to establish the exact mapping between halos and galaxies.


The first test case $\rm T_{TNG}$ relies only on the TNG data. 
It is conducted separately for both   
the star-forming sample, $\rm S=S_{SF}$, and the quenched sample, 
$\rm S=S_{Q}$ (see \S\ref{ssec:sample} and Table~\ref{tab:def-samples} for sample 
definitions). To test the performance of the model, we randomly 
split each of the TNG samples, $\rm S$, into a training set 
and a test set, with a ratio of $3:1$ in the number of 
galaxies between them. The steps are the following: 
\bit 
	\item Following procedure $P_1$, we apply the PCA to the histories $\tilde{\bfrm[h]}_{\rm halo}$ of the hosting halos of galaxies in 
	sample $\rm S$, which gives the transformation, 
	$\mathbb{T}_{\bfrm[e]_{\rm halo},\bfrm[o]_{\rm halo}}$, and 
	the low-dimension representation of the halo MAH, 
	$\bfrm[y]_{\rm halo}$.
	
	\item We apply the PCA to the SFH, $\tilde{\bfrm[h]}_*$, 
	of the galaxies in sample $\rm S$, which gives the transformation, $\mathbb{T}_{\bfrm[e]_*,\,\bfrm[o]_*}$, and the low-dimension 
	representation of the SFH, $\bfrm[y]_*$.
	
	\item Using the training set, we train the GBDT regressor, $\mathbb{R}$, 
	which maps $(\bfrm[y]_{\rm halo},\,\bfrm[x]_{\rm halo})$ into $(\bfrm[y]_*,\,\bfrm[x]_*)$.
	
	\item We apply the transformations obtained above to map halo properties 
	to galaxy properties in the test set using $(\bfrm[h]_*,\, \bfrm[x]_*) = \mathbb{T}^{-1}_{\bfrm[e]_*,\,\bfrm[o]_*}\,\mathbb{R}\,\mathbb{T}_{\bfrm[e]_{\rm halo},\bfrm[o]_{\rm halo}}(\bfrm[h]_{\rm halo},\,\bfrm[x]_{\rm halo})$, 
	and perform the post-processing.
\eit 
After these steps, we obtain the modeled SFR and $\MstarInt$ histories 
for the galaxies in the test set, and we compare them with the TNG 
data. Because the separation of the star-forming and quenched 
galaxies and all of the transformations are obtained directly 
from the simulation data, the performance of this test case can be 
viewed as the upper limit of our model. In this case, the deviation of 
the model output from the simulation is due to the intrinsic 
incapability of the model, which, in principle, can be improved 
by including more halo properties into $\xhalo$ and using more 
PCs of $\hhalo$, provided that the training set is sufficiently large.

The second test case $\rm T_{EAGLE^{+TNG-modes}}$ relies both on the TNG 
and EAGLE and is designed to mimic the situation in 
real applications where some of the model ingredients are unknown. 
The test is also made for both the star-forming 
sample $\rm S = S_{SF}$ and the quenched sample $\rm S = S_{Q}$ in EAGLE (see \S\ref{ssec:sample} and Table~\ref{tab:def-samples}). To test the model performance, we randomly split 
each of the EAGLE sample, $\rm S$, into a training set and a test set, 
again with a $3:1$ ratio in the number of galaxies between the two sets.
The test is conducted through the following steps:
\bit 
	\item In a real application, halo information is accessible. 
	Therefore, we directly apply the PCA to the histories, 
	$\tilde{\bfrm[h]}_{\rm halo}$, of EAGLE halos in sample $\rm S$, 
	which gives us the transformation, 
	$\mathbb{T}_{\bfrm[e]_{\rm halo},\bfrm[o]_{\rm halo}}$ 
	and the PCs describing the halo MAH.
	
	\item SFH is not accessible because it is the target of the model. 
	This prevents us from getting a dimension reduction template $(\bfrm[e]_*,\,\bfrm[o]_*)$. Thus, some assumptions have to be made. 
	We choose to use the eigenvectors, $\bfrm[e]_*$, that are built 
	from the TNG in $\rm T_{TNG}$, and we interpolate 
	each of these eigenvectors to the redshifts of EAGLE's snapshots. 
	In doing so, we in effect borrow the template from the TNG 
	for the analysis of EAGLE. As we will show later, using the 
	TNG template to reduce the dimension of EAGLE SFH   
	is, in terms of model performances, comparable to using the 
	template from EAGLE itself. Thus, only $\bfrm[o]_*$ remains to 
	be modeled in real applications, 
	and it can be modeled by using some parametric form to be 
	constrained by observations. As more observations are added, 
	the estimate of $\bfrm[o]_*$ will be improved. Here, we want to test 
	the upper performance limit of our model by using the real 
	$\bfrm[o]_*$ obtained directly from sample $\rm S$ of EAGLE, 
	and we denote it by $\tilde{\bfrm[o]}_*$. 
	Finally we obtain the 
	transformation, $\mathbb{T}_{\bfrm[e]_*,\,\tilde{\bfrm[o]}_*}$.
	
	\item In a real application, the mapping, $\mathbb{R}$,	also needs 
	to be modeled and constrained by observations. Again, 
	because we want to gauge the upper limit of the model performance, 
	we train $\mathbb{R}$ by halo properties, $\mathbb{T}_{\bfrm[e]_{\rm halo},\bfrm[o]_{\rm halo}}(\bfrm[h]_{\rm halo},\,\bfrm[x]_{\rm halo})$, 
	and galaxy properties, $\mathbb{T}_{\bfrm[e]_*,\,\tilde{\bfrm[o]}_*}(\bfrm[h]_*,\,\bfrm[x]_*)$, 
	both from the training set of EAGLE. The trained regressor is denoted 
	by $\tilde{\mathbb{R}}$.
	
	\item We apply the transformation $\mathbb{T}^{-1}_{\bfrm[e]_*,\,\tilde{\bfrm[o]}_*}\,\tilde{\mathbb{R}}\,\mathbb{T}_{\bfrm[e]_{\rm halo},\bfrm[o]_{\rm halo}}$ to the host halos of the test galaxies, and perform the post-processing to get the final output.
\eit 
In the end of all these steps, we obtain the model predictions 
for galaxy properties and compare them with the results of EAGLE.

Since in $\rm T_{EAGLE^{+TNG-modes}}$ some of the transformation 
ingredients are borrowed from TNG, the model performance is 
inevitably worse than that using the true transformation. 
To see the effect caused by the imperfect transformation, we design 
a third test case, $\rm T_{EAGLE}$, which is identical to $\rm T_{TNG}$, 
except that both $\rm S_{SF}$ and $\rm S_{Q}$ are taken from EAGLE.

Finally, we design a more realistic testing case, $\rm T_{join}$, 
in which the separation of star-forming and quenched galaxies 
is also to be modeled. This test is conducted for both the 
star-forming sample $\rm S=S'_{\rm SF}$ and the quenched sample 
$\rm S=S_{\rm Q}$ in EAGLE. We again use a 3:1 split 
between the training and test sets. The testing steps are the following: 
\bit 
	\item We apply the same modeling as in $\rm T_{EAGLE^{+TNG-modes}}$ 
	to $\rm S'_{SF}$ and $\rm S_Q$, and obtain two models that map 
	halo properties to the galaxy SFH separately for star-forming and 
	quenched galaxies.

	\item Using the combination of the training set in 
	$\rm S'_{SF}$ and $\rm S_Q$, we train a GBDT classifier which 
	classifies a $z=0$ galaxy into the star-forming or the 
	quenched population according to its halo properties, $\vmax$ at $z=0$,
	$z_{\rm infall}$, $z_{\rm mb,\, core}$ and the first three 
	PCs of the $\vmax$ history. 
	The inclusion of $z_{\rm infall}$ and $z_{\rm mb,\,core}$ is 
	motivated by the fact that these two properties are important 
	in affecting galaxy quenching (see \S\ref{ssec:reason-to-quench}). 
	
	\item We apply the classifier to the combination of the test sets 
	in both $\rm S'_{SF}$ and $\rm S_Q$. A galaxy is then classified 
	either as star-forming or quenched. We apply the two trained 
	models to star-forming and quenched populations, respectively.
\eit 

As mentioned in \S\ref{ssec:reason-to-quench}, the separation of 
star-forming and quenched galaxies is far from perfect, which can lead to
significant contamination in both the star-forming and 
quenched samples classified. This limits the performance of  
models based on halo properties. However, as 
we will show later, although the reconstruction of the SFH for 
individual galaxies is contaminated by imperfect classification, 
the statistical properties of the whole population are unbiased.
The final outputs of the two models in this test case consist of
properties of both star-forming and quenched galaxies at $z=0$, 
and are compared to the EAGLE data. As described above,
the separation of star-forming and quenched galaxies, as well 
as the transformation, all mimic real applications in this case.

We again want to see if the use of the template from 
EAGLE itself can make an improvement in the model performance. 
To this end, we define a fifth test case, $\rm T'_{join}$, 
which is identical to $\rm T_{join}$, except that 
the dimension reduction template is from EAGLE itself
in its first step.

\subsection{The results}
\label{ssec:model-result}

We now show the results of the five test cases, 
$\rm T_{TNG}$, $\rm T_{EAGLE^{+TNG-modes}}$, $\rm T_{EAGLE}$,
$\rm T_{join}$, $\rm T'_{join}$.
In the first three cases, the star-forming sample 
$\rm S_{SF}$ and the quenched sample $\rm S_{Q}$ are modeled separately. 
The modeled SFHs of star-forming galaxies, represented by 
SFR and $\MstarInt$ at each snapshot, are shown in Fig.~\ref{fig:sf-model-history} 
in comparison with the simulated one.
The simulated SFR histories of individual galaxies show small fluctuations 
on small time scale, which are not captured well by the PCs in the 
model, but can be modeled by adding a random component in the 
post-processing. In all of the three test cases, the model successfully 
reproduces the overall trend of the simulated SFR histories for individual galaxies. 
The difference between the model and the simulation is small at low 
$z$, and becomes slightly larger at higher $z$ where the SFR becomes 
too low to model accurately.
The modeled SFR histories in $\rm T_{EAGLE^{+TNG-modes}}$ is as 
good as those in $\rm T_{TNG}$ and $\rm T_{EAGLE}$ at low redshift, 
and  becomes slightly worse at high $z$ in some cases.
The $\MstarInt$ histories are more smooth, but the overall conclusion 
for the SFR histories also holds for the $\MstarInt$ histories.
All of these indicate that the model can reproduce both the 
SFR and $\MstarInt$ histories for star-forming galaxies. 

To quantify the goodness of the model in describing the data of
star-forming galaxies, we compare 
in Fig.~\ref{fig:sf-model-pp-compare} the simulated SFR 
and $\MstarInt$ at several redshifts from $0$ to $3$. 
We also compute the mean and standard deviation of the 
$\Mstar$-weighted average of residual between the simulation and the model.
The results can be summarized as follows. 
(i) The residual between modeled and simulated 
$\log\,{\rm SFR}$ and $\log\,\MstarInt$ has no obvious bias at all redshifts. 
(ii) The residual between modeled and simulated SFR and $\MstarInt$ 
slightly increases with redshift. The scatter is about 
$0.3\, {\rm dex}$ for the SFR and $0.1\, {\rm dex}$ for the $\MstarInt$ 
at $z=0 \sim 2$ and increases at $z> 2$.
(iii) The random noise, which cannot be modeled by the 
halo MAH, is moderate at low-z, and becomes 
significant at $z=3$. 
Because numerical simulations usually have more limited 
output time resolution at higher redshift, the SFHs of 
galaxies at higher redshift are expected to contain more noise. 
These suggest that the full potential of the empirical model is limited
by the resolution and output frequency of the hydrodynamic simulation used.
Consistent with this, the results in \S\ref{ssec:relations-in-main-seq}
show that the residual in the sSFR cannot be fully explained by halo properties.
(iv) The bias and scatter for both SFR and $\MstarInt$ at all 
redshifts are only slightly larger in $\rm T_{EAGLE^{+TNG-modes}}$ 
than in $\rm T_{EAGLE}$, indicating that the borrow of template does 
not introduce large error in the model.
All these confirm that the model is powerful in describing 
the SFH of star-forming galaxies.

The modeled SFHs obtained from the quenched sample 
$\rm S_Q$ in the first three test cases 
are shown in Fig.~\ref{fig:qch-model-history} in comparison with 
the simulation results. Compared with the results for 
star-forming population, the SFR and $\MstarInt$ histories 
are as well reproduced over a wide range of redshift. 
Case $\rm T_{EAGLE^{+TNG-modes}}$, which uses the TNG template,  
also gives results comparable to cases where EAGLE template 
itself is used. The only exception is at low redshift when these 
galaxies are quenched and the SFRs decrease quickly to very low 
values for the model to predict accurately. 
However, even in this case, the predicted $\MstarInt$ histories still 
closely follow the simulated ones.

For the quenched sample $\rm S_Q$, we also show, in
Fig~\ref{fig:qch-model-pp-compare}, the comparisons 
between the model predictions and the simulated results for both 
SFR and $\MstarInt$ at several redshifts between $0$ to $3$. 
At low redshift, the SFR of quenched galaxies cannot be predicted 
accurately, so that both the bias and scatter are large. 
As we go to higher redshift, the bias and scatter decrease. 
These indicate that, even for a galaxy that is quenched at 
$z=0$, it is still possible to infer its SFH from its halo MAH. 
In all of the three test cases, the modeled $\MstarInt$ is tightly 
correlated with the simulation results, with almost no bias 
at low $z$ and small bias at $z=3$, and with small scatter at all 
redshifts. Again, the use of TNG template to model EAGLE galaxies 
in $\rm T_{EAGLE^{+TNG-modes}}$ is as good as using EAGLE's own template
in $\rm T_{EAGLE}$, indicating that the model can reproduce the SFH
even for quenched galaxies.

Based on these test results from $\rm T_{TNG}$, $\rm T_{EAGLE^{+TNG-modes}}$ and $\rm 
T_{EAGLE}$, we conclude that our model can describe accurately the SFH of both 
star-forming and quenched galaxies except for the current SFR
of quenched galaxies. Thus, if we can find a way to separate the 
star-forming and quenched populations, a model can be constructed for 
both populations. In the following, we show that a statistically 
correct model can be constructed even if a clean separation 
between the two population is not feasible 
(because of the reason discussed in \S\ref{ssec:reason-to-quench}).

\begin{figure*}\centering
	\includegraphics[width=15cm]{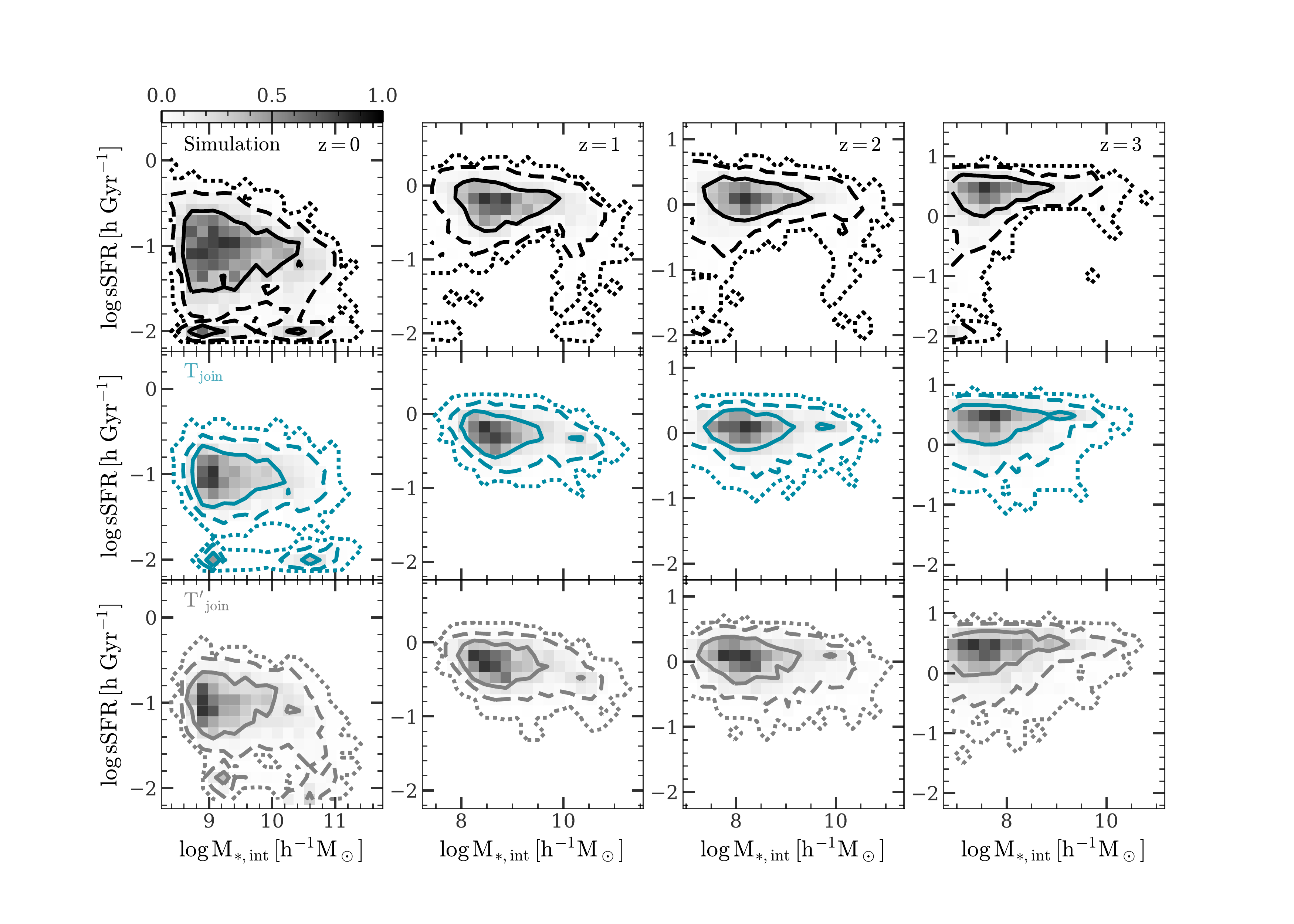}
	\caption{
	  The relation between the smoothed sSFR and $\MstarInt$ in the
          histories of all the test galaxies in
          $\rm T_{join}$ and $\rm T'_{join}$, in comparison with
          the EAGLE simulation. Panels from top to bottom are
          for the EAGLE simulation, cases $\rm T_{join}$ and $\rm T'_{join}$, 
          respectively.
	      Each column shows the relation at a specific redshift
          as indicated in the first row. {\bf \color{Gray} Gray}
          shades are normalized histograms.
	}
	\label{fig:mix-model-ms-vs-ssfr}
\end{figure*}

\begin{figure*}\centering
	\includegraphics[width=15cm]{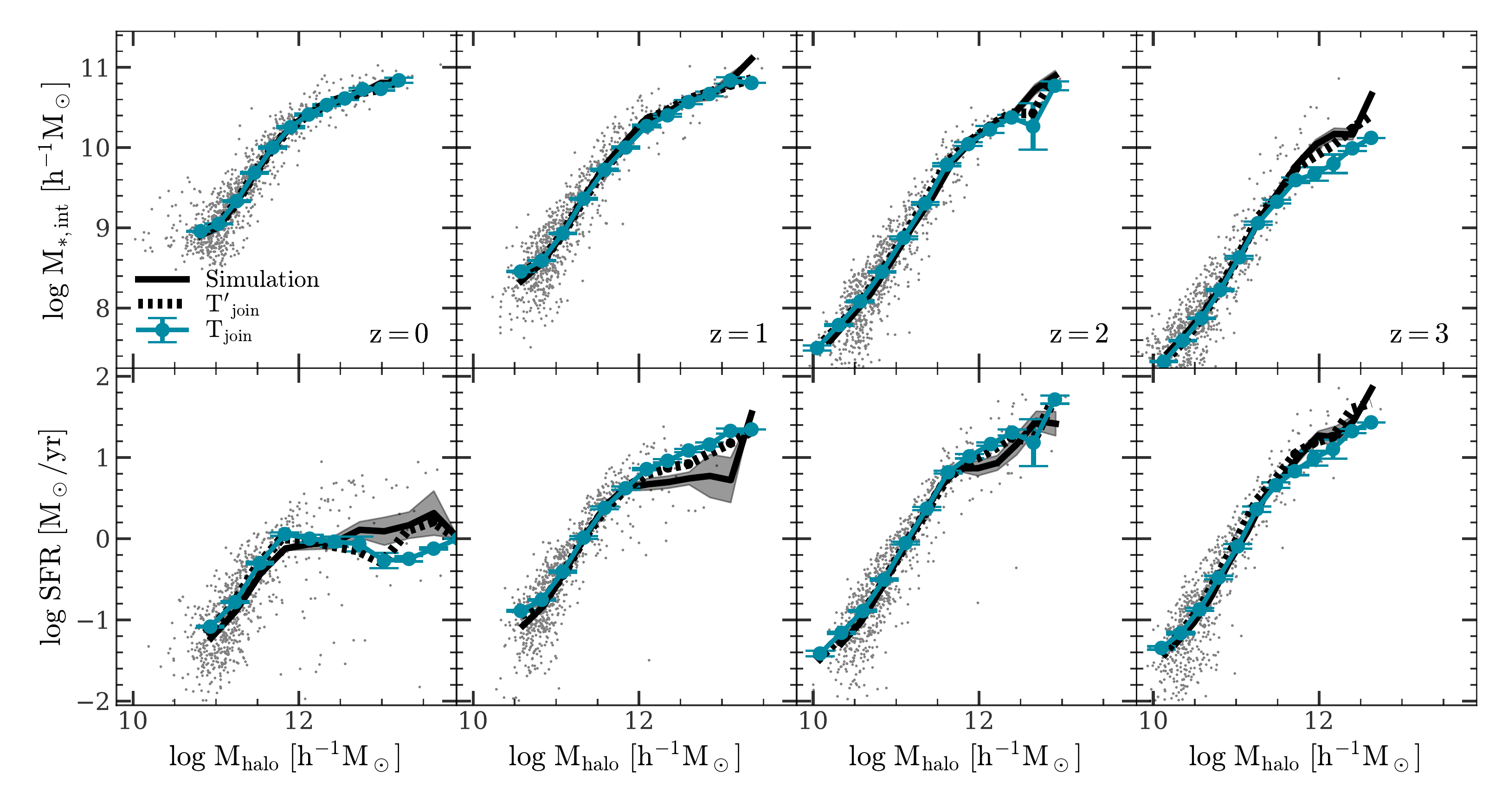}
	\caption{
	  The relation between halo mass $\Mhalo$ and stellar properties
          ({\bf upper} row: $\MstarInt$; {\bf lower} row: the smoothed SFR)
          at different redshifts (as indicated) in the histories of
          all the test galaxies in $\rm T_{join}$
          ({\bf \color{TealBlue} green} line with error bars indicating
          the standard deviation) and $\rm T'_{join}$ ({\bf dashed black} line).
          In each panel, {\bf \color{Gray} gray} dots are from the
          EAGLE simulation; the {\bf solid black} line and gray shade
          indicate the mean and standard deviation, respectively. 
          Galaxies with $\ssfr > 10^{-2}\gyri$ are used for the SFR results.
	}
	\label{fig:model-stellar-vs-halo}
\end{figure*}

\begin{figure}\centering
	\includegraphics[width=\columnwidth]{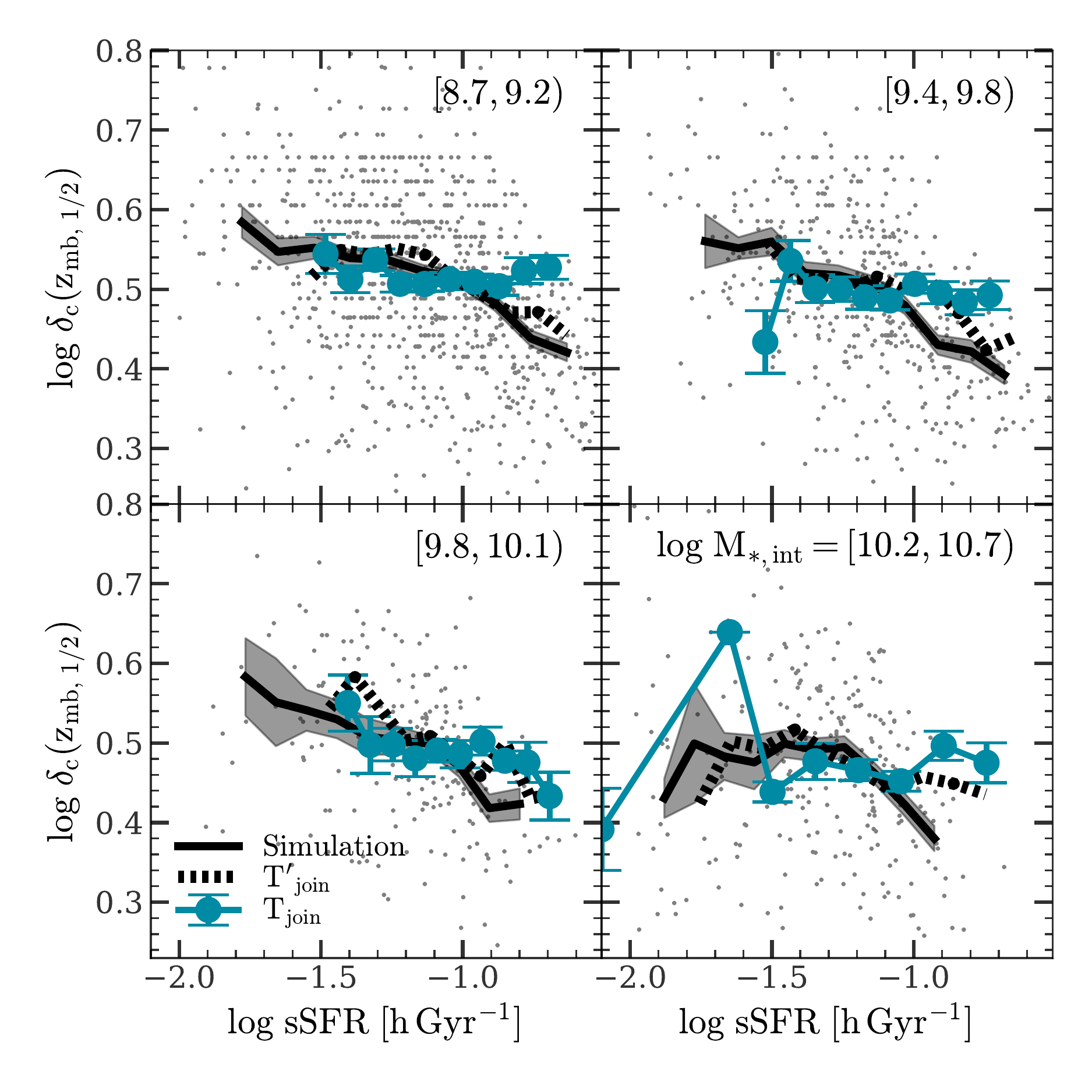}
	\caption{
	  The relation between halo half-mass formation time,
          $z_{\rm mb,\,1/2}$ and galaxy sSFR for $z=0$ galaxies
          with different $\MstarInt$, as indicated in each panel.
          In each panel,
          {\bf \color{Gray} gray} dots are from the EAGLE simulation,
          with the {\bf solid black} line and shade indicating
          the mean and standard deviation of the mean, respectively. 
          Test results using $\rm T_{join}$ are shown by 
          the {\bf \color{TealBlue} green} line with error bars,
          while those using $\rm T'_{join}$ are shown by the 
          {\bf dashed black} line. Only galaxies with
          $\ssfr > 10^{-2}\gyri$ are used. 
	}
	\label{fig:model-mbhalf-vs-ssfr}
\end{figure}

In the remaining two test cases, we need to first classify a galaxy 
as star-forming or quenched, and then model it by the trained model
appropriate for its class. Fig.~\ref{fig:mix-model-ms-vs-ssfr}
shows the results based on $\rm T_{join}$, where the distributions 
of model galaxies in the $(\log\, \MstarInt,\ \log\, \ssfr)$ plane
at four different redshifts are compared with the simulated results. 
At $z=0$, model galaxies show a bimodal distribution, 
consistent with the simulation results. At higher redshifts, the 
simulation shows some weak sign of bimodality, which is
not well captured by the model. In the simulation, the mean 
value of $\ssfr$ of the main sequence increases slowly 
with redshift, a trend that is well reproduced by the model. 
Consistent with the simulation, the scatter in 
the modeled main sequence decreases with redshift. 
However, the predicted amount of scatter at $z=0$ is 
smaller than that in the simulation, which is due to the 
limited degrees of freedom of the random component used 
in the post-processing. 
Fig.~\ref{fig:mix-model-ms-vs-ssfr} also
shows the galaxy distribution using $\rm T'_{join}$ and
EAGLE's own templates to model EAGLE galaxies. 
It is clear that the use of TNG templates in $\rm T_{join}$ is as 
good as using EAGLE's own template in $\rm T'_{join}$.

In Fig.~\ref{fig:model-stellar-vs-halo}, we show the ${\rm SFR} - \Mhalo$ 
and $\Mstar - \Mhalo$ relations at four redshifts from $z=0$ to $z=3$
for the test case $\rm T_{join}$, in comparison with the simulation
results. For comparison, we also show the results from the test case 
$\rm T'_{join}$ to test the effect of borrowing external template. 
As one can see, the $\Mstar - \Mhalo$ relations predicted by the model 
for $\rm T_{join}$ match well the simulation results. Only at $z>2$ 
is the modeled $\Mstar-\Mhalo$ relation slightly lower. 
Compared with the results for $\rm T'_{join}$, which match the simulation 
results almost perfectly, this small difference
is clearly produced by the use of the imperfect template
in $\rm T_{join}$.

The modeled ${\rm SFR}-\Mhalo$ relation in $\rm T_{join}$ is also 
similar to that in the simulation, with moderate discrepancy 
at $\Mhalo > 10^{12}\msun$. Comparing this to the predictions of 
$\rm T'_{join}$, which match the simulation results better but 
not perfectly, we infer that this discrepancy is partly due to 
the use of imperfect template in $\rm T_{join}$ and partly due to 
the imperfect classification of star-forming and quenched galaxies
in both cases. As discussed in \S\ref{ssec:relations-in-quench},
the decision boundary is ambiguous for high-mass galaxies, 
which are hosted by massive halos, and 
a slightly offset in the decision is likely to produce a 
significantly different result.

Since we have already included PCs of $\vmax$ as features in 
the regressors, our model is expected to reproduce the 
dependency of galaxy properties on halo MAH, a phenomenon 
usually referred to as the ``assembly bias''. 
To demonstrate this, we plot the relation between the 
halo half-mass formation time, $z_{\rm mb,\,1/2}$, and galaxy 
sSFR for galaxies at $z=0$ for case $\rm T_{\rm join}$. 
The results for four different stellar mass bins are shown in 
Fig.~\ref{fig:model-mbhalf-vs-ssfr}, where we also include the 
results from the simulation and from $\rm T'_{join}$ for 
comparison. In all of the mass bins, galaxies in halos of
earlier assembly on average have smaller sSFR. Both $\rm T_{join}$ 
and $\rm T'_{join}$ can reproduce this trend. 

The results of $\rm T_{join}$ have a small bias relative to the simulation. 
Because the effects of halo PCs on galaxy SFH are 
much smaller than the total scatter of the star-forming 
main sequence (see \S\ref{ssec:relations-in-main-seq}), 
the regressor that maps halo PCs to galaxy properties tends to 
reduce the model variance at the cost of increasing bias. 
When the SFH template adopted in the model is imperfect, 
the bias is larger, as is seen in the results of $\rm T_{join}$
in comparison with those of $\rm T'_{join}$. 
Overall, our model reproduce correctly the assembly bias
in the data, especially when the PC template can 
account for the variance in the SFH of galaxies. 

To conclude, the tests using $\rm T_{join}$ and $\rm T'_{join}$ 
demonstrate that our empirical model can describe the 
galaxy-halo relation correctly in a statistical way, 
even though the classification between star-forming and  
quenched galaxies is not accurate for individual galaxies.

\section{Summary}
\label{sec:summary}

In this paper, we use the TNG and EAGLE simulation data to infer the galaxy-halo 
relations that are needed to build an empirical model for central galaxies in 
dark matter halos. Our analysis is based on PCA for dimension reduction and 
GBDT for regression and classification. Our main results and their implications 
are summarized as follows.
\begin{enumerate}[leftmargin=1em,itemsep=0pt,parsep=0pt,topsep=0pt]
	\item The star-forming main sequence is a well-defined population driven by $\vmax$ of host halos. The $\Mstar$-$\vmax$ and SFR-$\vmax$ relations 
	for this population at $z=0$ are both tight, with $R^2 \ge 0.9$ and $0.7$, respectively, and they are even tighter at higher $z$. 
	Other halo properties are secondary and provide only small improvements 
	in the predictions of $\Mstar$ and SFR.
	
	\item The residual of the SFR-$M_*$ relation for the main sequence, 
	represented by $\ssfrRes$, is not dominated by any halo property 
	tested in this paper. Using a combination of a large set of halo 
	properties, the value of $R^2$ in the prediction of $\ssfrRes$ is 
	still $<0.5$ at both low and high $z$. These indicate that 
	modeling the SFR based on halo properties with the use 
	of deterministic relation between the two can lead to 
	spurious and biased results. A random component is needed in 
	order to model SFR in a statistically unbiased way.
	
	\item The quenching of a low-mass central galaxy is tightly correlated 
	with the infall-ejection process of the host halo. In contrast, 
	the quenching of a high-mass central galaxy is related closely 
	to the formation of a massive progenitor in its host halo at high $z$,
	as indicated by the core formation redshift, $z_{\rm mb,\,core}$. 
	For both low-mass and high-mass galaxies, it is difficult to 
	train classifiers that can separate the star-forming from 
	the quenched population, because of the sample imbalance and 
	overlapped distribution between these two populations.

	\item For the quenched population,  $\MstarInt$ is tightly 
	correlated with halo $\vmax$. The $\MstarInt$ at $z=0$ depends 
	predominantly on $\vmax$, while PCs of the $\MstarInt$ history 
	are correlated with the PCs of the $\vmax$ history. In general, 
	the higher order PCs of $\MstarInt$ are less well recovered 
	by the regressors.
	
\end{enumerate}

Based on the inferred galaxy-halo relations, we propose an empirical model 
for star formation in central galaxies of dark matter halos. The main procedures can be 
summarized as follows.
\begin{enumerate}[leftmargin=1em,itemsep=0pt,parsep=0pt,topsep=0pt]
	\item The empirical model consists of three procedures, which 
	reduce the dimension of halo MAH by the PCA, map the halo properties 
	into stellar properties by the GBDTs, and recover the dimension of 
	the SFH by the inverse of the PCA.
	
	\item For both star-forming and quenched galaxies, the 
	empirical model shows good performances in all of our test cases. 
	The reconstructed SFHs of individual galaxies follow the correct 
	trends in comparison with the simulated results. 
	The SFR and $\MstarInt$ at all redshifts are reconstructed with 
	small bias and small residuals. The only exception occurs for some 
	quenched galaxies where the SFRs in the simulations decrease 
	too rapidly to capture by the model.  
	
	\item Central galaxies can be classified into star-forming and quenched 
	populations on the basis of halo properties, and can be modeled
	separately according to their classes. Although the classification 
	is imperfect and has contamination between the two classes, 
	the predicted statistical properties of the galaxies match   
	well with the simulation inputs. These include the 
	bimodal distribution of galaxies in the SFR - stellar mass
        diagram, the stellar mass - halo mass 
	and SFR - halo mass relations of galaxies at different $z$,  and 
	the assembly bias of galaxies.
\end{enumerate}

The results presented here provide a framework of using hydrodynamic 
simulations to discover ingredients that can be included in empirical 
models of galaxy formation and to build templates that 
can be used to reduce the model complexity.
In the future, we will extend our analysis by including satellite 
galaxies. The results obtained in this paper can be used as 
the initial conditions before a galaxy becomes a satellite,
and the subsequent evolution of the satellite population
is to be modeled again on the basis of halo properties, such as 
halo masses and merging orbits. With these, we will 
build a full empirical model based on the architecture 
provided by numerical simulations.  

\section*{Acknowledgements}
This work is supported by the National Key R\&D Program of China
(grant No. 2018YFA0404502), and the National Science 
Foundation of China (grant Nos. 11821303, 11973030, 11673015, 
11733004, 11761131004, 11761141012). We acknowledge the Virgo Consortium for making their EAGLE simulation data available.
We acknowledge Dandan Xu, Yuning Zhang and Jingjing Shi for accessing the TNG simulation data.
Y.C. and K.W. gratefully acknowledge
the financial support from China Scholarship Council.

\section*{Data availability}
The data and software underlying this article will be shared on reasonable 
request to the corresponding author. They are available at
\url{https://www.chenyangyao.com/publication/20/empirical-model-satellite/}. 
The computation in this work is supported by the HPC toolkit HIPP at \url{https://github.com/ChenYangyao/hipp}.

\bibliographystyle{mnras}
\bibliography{references}

\appendix

\section{PCA of galaxy and halo formation histories}
\label{app:PCA}

\begin{figure}\centering
	\includegraphics[width=\columnwidth]{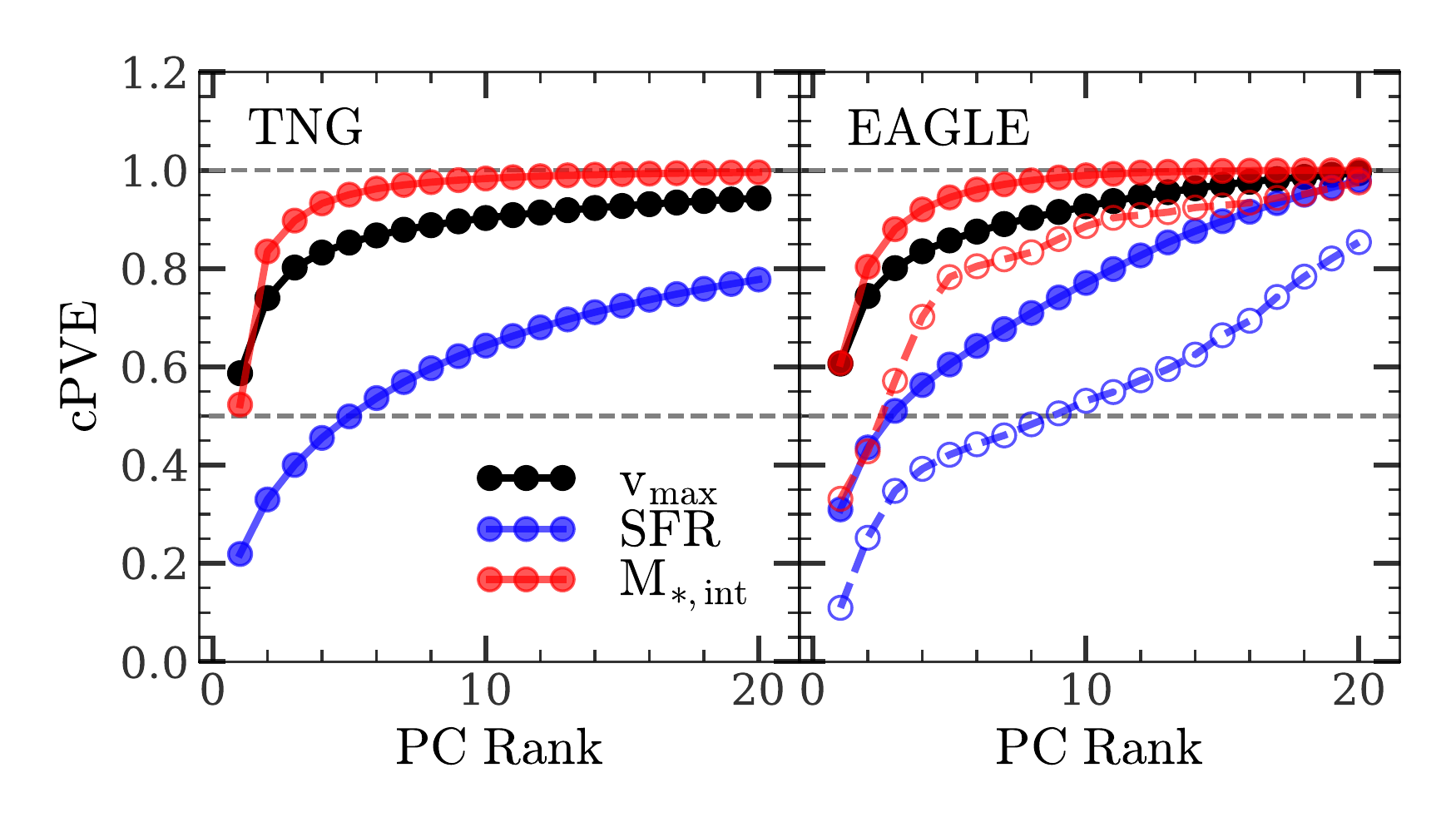}
	\caption{
		cPVE curves for $\vmax$, SFR and $\MstarInt$ histories of halos and galaxies in TNG ({\bf left}) and EAGLE ({\bf right}). In the right panel, open circles joined by dashed lines are cPVE curves for EAGLE SFR ({\bf \color{Blue} blue}) and $\MstarInt$ ({\bf \color{Red} red}) histories using PCA templates from TNG.
	}
	\label{fig:cpve}
\end{figure}

\begin{figure}\centering
	\includegraphics[width=\columnwidth]{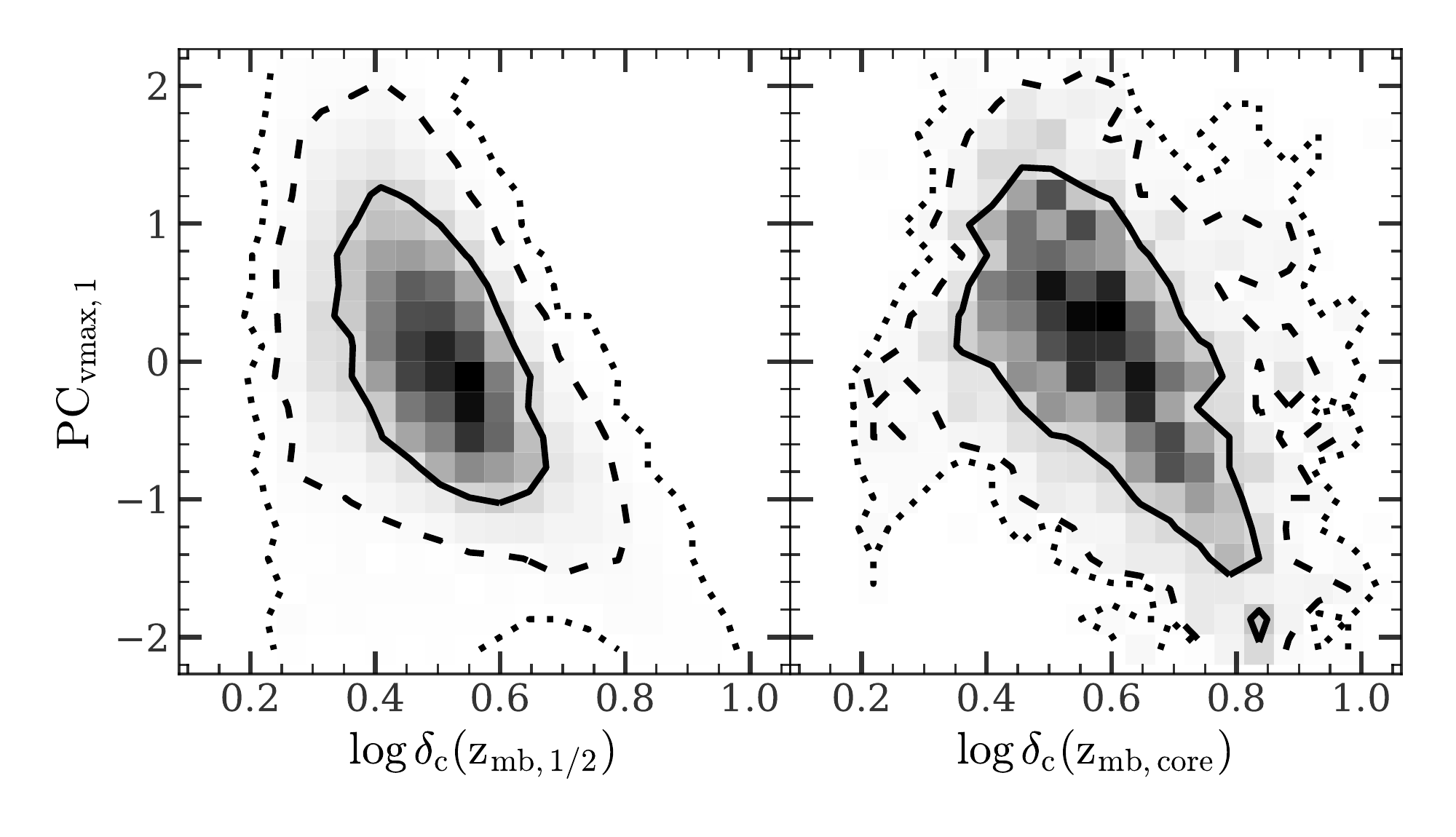}
	\caption{
		Relation between the first PC of halo $\bfrm[v]_{\rm max}$ history and formation time $z_{\rm mb,\,1/2}$ ({\bf left} panel; using all $z=0$ TNG halos with $\Mhalo \ge 10^{8.5} \msun$) and $z_{\rm mb,\,core}$ ({\bf right} panel; using all $z=0$ TNG halos with $\Mhalo \ge 10^{10} \msun$, because small halos do not have $z_{\rm mb,\,core}$ measurements).
	}
	\label{fig:pc-vs-ftimes}
\end{figure}

\begin{figure*}\centering
	\includegraphics[width=15cm]{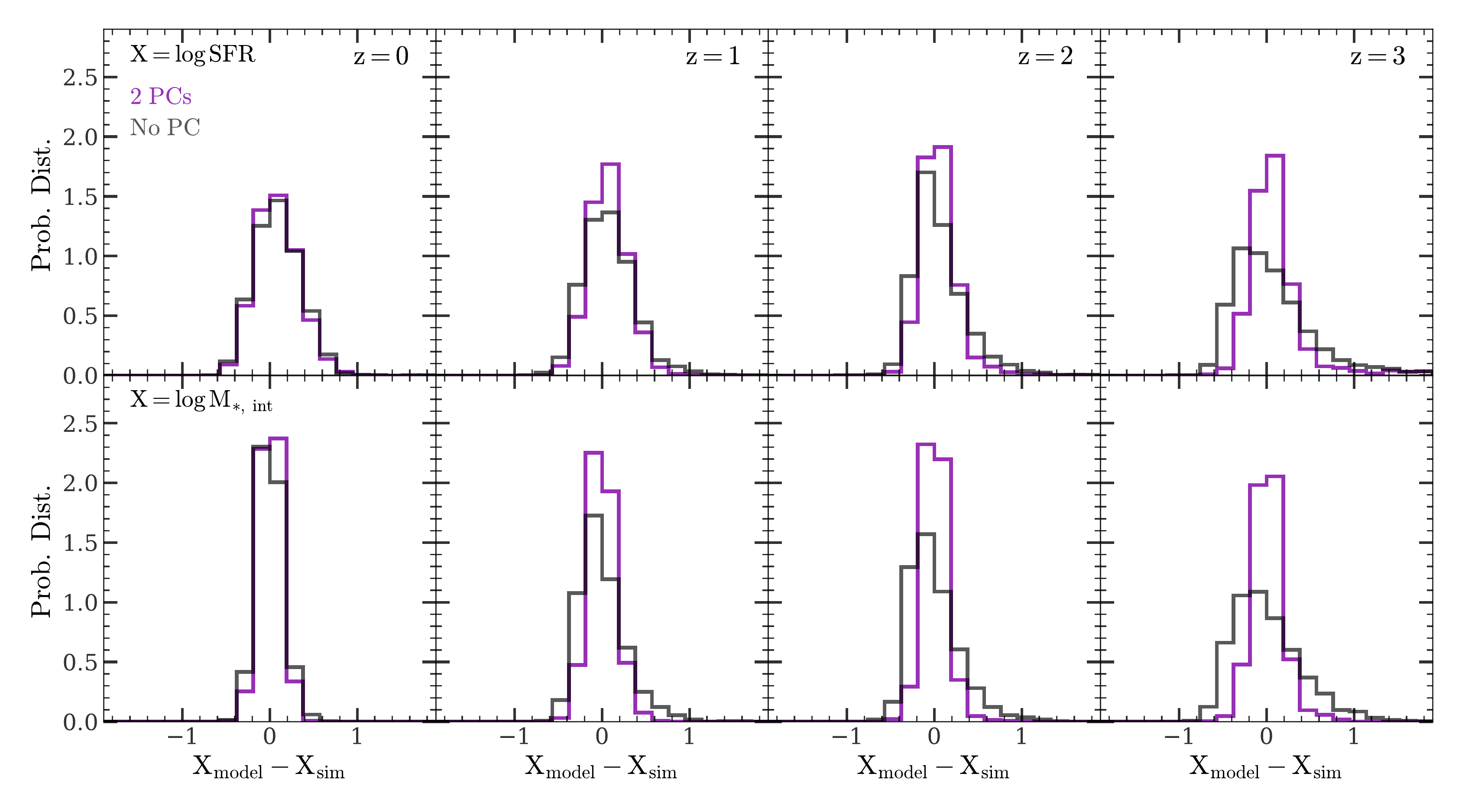}
	\caption{
		Distributions of the $\log\,{\rm SFR}$ ({\bf upper row}) and $\log\,\MstarInt$ ({\bf lower row}) difference between model galaxies and TNG simulated galaxies. Here we show the test galaxies in the sample $S_{\rm SF}$ in the first test case $\rm T_{TNG}$. {\bf \color{Purple} Purple} histograms are from the model using the first two PCs for MAH and SFH. {\bf Black} histograms are from the model without any PC.
	}
	\label{fig:pc-to-noPC}
\end{figure*}

The Principal Component Analysis (PCA) is an unsupervised, 
reduced linear Gaussian dimension-reduction method \citep{PearsonKarl:1901:PCA,HotellingH:1933:PCA}. As demonstrated in \cite{ChenYangyao:2020:Halo-structure}, the halo mass assembly history
(MAH), which is a vector in high-dimensional space, can be effectively 
reduced to several Principal Components (PCs) that still capture 
most of the sample variance. Here we briefly describe how we 
apply the PCA to galaxy and halo formation histories. 
A modern and detailed theoretical description of the PCA 
can be found in \cite{BishopC:2006:PRML}.

The various ``history'' quantities considered in this paper
are also vectors in high dimension space, and we use the 
PCA to reduce their dimensions so that each history can 
be described by a set of PCs. For each of the histories, $\bfrm[h]$ 
($\bfrm[h] = \bfrm[v]_{\rm vmax}$, $\bfrm[SFR]$ or $\bfrm[M]_{\rm *,\,int}$), 
we apply the PCA according to the following steps.
\bit 
\item Because of the resolution limit of the simulations, 
the history of a galaxy cannot be traced back to an arbitrarily 
high redshift. For a galaxy sample $\rm S$, we trim $\bfrm[h]$ 
of each galaxy above a chosen redshift, so that $90\%$ of the 
galaxies have history measurements for the remaining redshifts. 
Galaxies that do not have history measurements at some of 
the remaining redshifts are padded with a small value 
to ensure numerical stability.

\item We make a proper transformation of $\bfrm[h]$ according 
to the description given in \S\ref{ssec:model} to make it suitable 
for PCA. The transformed history is denoted by $\tilde{\bfrm[h]}$.

\item We apply the PCA to $\tilde{\bfrm[h]}$ of all galaxies in $\rm S$. 
The PCA gives a mean offset $\bfrm[o]$, and a set of new base 
vectors $\bfrm[e]_{i} (i=1,2,3,...)$ whose eigen-values, 
$\lambda_i (i=1,2,3,...)$, are ranked in a descending order. 
The history is then transformed into the new frame by
\begin{equation}
\bfrm[PC] = (\bfrm[e]_1,\,\bfrm[e]_2,\,...)^{\rm T} (\tilde{\bfrm[h]}-\bfrm[o]).
\end{equation} 
\eit 
To reduce the dimension of $\tilde{\bfrm[h]}$, we can keep 
a set of $m$ important PCs, 
$\bfrm[PC]_m=({\rm PC}_1, {\rm PC}_2, ..., {\rm PC}_m)^{\rm T}$. 
We can reconstruct $\tilde{\bfrm[h]}$ from $\bfrm[PC]_m$ using  
\begin{equation}
\tilde{\bfrm[h]}_{{\rm recon}, m}=(\bfrm[e]_1,\bfrm[e]_2,...,\bfrm[e]_m)\bfrm[PC]_m+\bfrm[o].
\end{equation}
This inevitably causes some loss of information. 
The information retained by $\bfrm[PC]_m$ is by described the 
cumulative proportional variance ratio (cPVE), defined as
\begin{equation}
{\rm cPVE}_m = \frac{{\rm Var}[\tilde{\bfrm[h]}_{{\rm recon},m}]}{{\rm Var}[\tilde{\bfrm[h]}]}.
\end{equation}
In \S\ref{ssec:model}, we consider a case mimicking 
real applications, in which the dimension-reduction templates from 
the TNG simulation are applied to reduce the dimension of the SFHs
of the EAGLE simulation. To this end, we first apply the PCA to 
both TNG and EAGLE. We then keep only the EAGLE offset vector $\bfrm[o]$, 
and replace all EAGLE base vectors $\bfrm[e]_i$ with the TNG base 
vectors interpolated to the redshifts of the EAGLE snapshots.  
Using this new frame, we can compute the PCs for each
EAGLE SFH, and measure the performance of the reconstruction 
by the corresponding cPVE. 

The cPVE as a function of $m$ is shown in Fig.~\ref{fig:cpve}. 
As one can see, when using the templates obtained from a simulation itself,  
the $\vmax$ and $\Mstar$ histories in both TNG and EAGLE converge 
quickly to $1$, indicating that the first several PCs 
take most of the variance. This shows that the main structures 
of the halo MAH and the stellar mass assembly history 
are fairly simple, and can be effectively described by 
a small number of parameters. For the SFR history,  
the first several PCs are still the most important ones, 
but cPVE increases slowly as $m$ increases, 
indicating that the SFR history is noisy on small time 
scales. This can be seen from the plots of SFR histories 
of individual galaxies presented in \S\ref{ssec:model-result}. 
It is thus only sensible to link the main structure of the 
SFR history to halo properties, but to treat the small scale 
fluctuation as a random (uncorrelated) component to be included 
in the empirical model. The design of our empirical model in 
\S\ref{ssec:model} exploits this idea. 

When using the TNG templates to describe EAGLE histories,  
the reconstruction is poorer, as shown by the open circles 
connected by dashed lines in Fig.~\ref{fig:cpve}. 
However, the first several PCs are still the most important ones 
and each of the higher order PCs contributes only a small 
fraction of the cPVE.

The parameterization using PCs for halo MAH has several advantages over
formation times.
PCs of MAHs are linearly orthogonal, therefore reducing the degeneracy. 
The first PC of
MAH has better correlation with halo concentration than other formation
times do, as shown in \cite{ChenYangyao:2020:Halo-structure}.  
PCs also have clear physical meaning. The first PC of MAH 
is tightly related to the formation time for halos with mass exceeding 
a certain value. Fig.~\ref{fig:pc-vs-ftimes} demonstrates its relation 
with two formation times, and we see a tight relation between them.
Higher order PCs naturally reflect more subtle properties in
the formation histories of halos, such as major mergers. 
\citep[see][for detailed discussions]{ChenYangyao:2020:Halo-structure}.

PCs of halo MAHs are also tightly correlated with the SFH of galaxies 
(see \S~\ref{sec:galaxy-halo-conn}).
Although $\Mhalo$ or $\vmax$ is the dominant factor in galaxy SFH, 
more information can be captured with the help of PCs, so that 
the empirical model is more powerful in describing the details of galaxy 
SFH. We show an example of this improvement in the Fig.~\ref{fig:pc-to-noPC}, 
where we use the test case $\rm T_{TNG}$ (see \S\ref{ssec:model-test}) to 
demonstrate the difference made by including the PCs of both MAH and SFH.

\section{Gradient boosted decision trees}
\label{app:GBDT}

Boosting is a large set of model ensemble methods that combine 
multiple weak learners (regressors or classifiers) to produce a strong 
learner capable of capturing complex patterns in statistical 
learning tasks. Compared with other ensemble methods, such as the 
random forest \citep{BreimanLeo:2001:RandomForest} that starts 
with strong learners and uses multiple-sourced randomness 
to suppress model variance, boosting methods are faster in 
computation and still maintain comparable performance.

A successful example of boosting methods is AdaBoost \citep{FreundY:SchapireRE:1997:AdaBoost}, which can be viewed 
as a ``greedy'' algorithm that optimizes an exponential objective 
function \citep[see, e.g.,][]{BishopC:2006:PRML}. The
extensions of this method to arbitrary differentiable objective 
functions can be made through gradient boosting or gradient boosted 
decision trees (GBDT) \citep{FriedmanJH:2001:GradientBoosting}, 
and stochastic optimization strategies \citep{FriedmanJH:2002:StochasticGradientBoosting}. 
The idea behind boosting motivates the developments of 
some modern deep neural networks, such as those with residual blocks 
\citep[ResNet, see][]{HeKaiming:2015:ResNet} and densely 
connected blocks \citep[DenseNet, see][]{HuangGao:2017:DenseNet}. 
In this paper, we use GBDT for both regression and classification.

The idea of boosting is to build a sequence of weak learners 
$f_i(\bfrm[x])\ (i=1,2,...,M)$, and combine them to form 
a regression function or classification function,
\begin{equation}
F_M(\bfrm[x]) = \sum_{i=0}^{M} f_i(\bfrm[x]).
\end{equation}
In regression problems, $F_M$ maps the feature variable $\bfrm[x]$ 
to the target value. In classification problems, $F_M$ maps 
$\bfrm[x]$ to the class probability, and the final prediction 
is chosen to be the class with the highest probability. 
Once we have a training data set $D=\{(\bfrm[x]_i, \bfrm[y]_i)\}_{i=1}^{N}$, 
the best $F_M$ is the one that minimizes the loss function $l(F_M|D)$.

Without any constraint, the optimization of $l$ is infeasible 
because the functional space of $f_i$ has infinity dimensions 
and the possible combinations are also infinite. The GBDT provides 
a tree-based ``greedy'' algorithm to solve this problem. 
Starting from an arbitrary naive learner $F_0$ (e.g., a constant function), 
the GBDT algorithm recursively adds new learner $f_M$ 
into $F_{M-1}$ to give $F_M=F_{M-1}+f_M$, such that 
$l(F_M|D) < l(F_{M-1}|D)$. To find the best $f_M$ at each 
iteration, we expand $l$ as a series,
\begin{equation}
l(F_M|D)=l(F_{M-1}|D)+f_M \cdot \nabla_F l(F|D) |_{F=F_{M-1}}.
\end{equation}
If $f_M$ is chosen such that $f_M=-\alpha \nabla_F l(F|D) |_{F=F_{M-1}}$, 
with $\alpha$ being the learning rate, then the loss function 
is guaranteed to decrease, and the iteration is an example of 
the gradient descent algorithm. In general, $f_M$ can be any function 
that is parallel with the gradient.

In our applications, we use only the loss function derived from the 
exponential family \citep[L2 loss in regression; cross-entropy loss in 
classification; see][]{BishopC:2006:PRML}, so that the gradient 
\begin{equation}
\nabla_F l(F|D)|_{F=F_{M-1}} = F_{M-1}(\bfrm[x]_i)-\bfrm[y]_i\ (i=1,2,...,N),
\end{equation}
which is the residual of $F_{M-1}$ relative to the real target 
values. With this choice, we train a shallow decision 
tree regressor $t(\bfrm[x])$ with the training set $\{(\bfrm[x]_i, F_{M-1}(\bfrm[x]_i)-\bfrm[y]_i)\}_{i=1}^{N}$, 
and finally obtain $f_M$ using $f_M=-\alpha t$. 
In the iterative process modeled above, trees built earlier 
mainly handle the large-scale structures in the feature space, 
while those built later focus on the local difficulties
that have not been captured.

The boosting algorithm defined above may have problems from 
over-fitting. To overcome them, we use the stochastic GBDT \citep{FriedmanJH:2002:StochasticGradientBoosting}. At each 
iteration step, we only use a random subset of the whole data set 
to train the tree $t(\bfrm[x])$. Such a randomness in the 
training set can effectively suppress the model variance, 
and is proved equivalent to ordinary regularization in 
some cases \citep[see, e.g.,][]{BishopCM:1995:Noise-eq-to-tikhonov}.

For our analysis, we use the \href{https://scikit-learn.org/stable/index.html}{scikit-learn} 
package to perform the GBDT. We choose the maximal depth of each 
tree to be three, which gives a sufficiently weak learner as 
required by boosting. A random subset of 75\% of 
the training data is used at each iteration step, which is sufficient to 
suppress over-fitting for most tasks. We adopt a small learning 
rate, $\alpha=0.08$, as recommended by \cite{HastieT:2001:ESL} 
to avoid overshot. We use $25\%$ of the training data as the 
validation set, and terminate the iteration if  
the validation performance is not improved in 10 
consecutive steps.

Once the ensemble of trees is built, the contribution of each 
variable $x \in \bfrm[x]$ in the prediction of target 
$\bfrm[y]$ can be described by an importance value, $\mathcal{I}(x)$.
This value is defined as the fraction of the decrease of the total 
loss caused by $x$ in the construction of each tree, satisfying the normalization
condition
\begin{equation}
	\sum_{x \in \bfrm[x]}\mathcal{I}(x) = 1.
\end{equation}
The definition of $\mathcal{I}(x)$ is motivated by the fact that the goal of 
a regressor or a classifier is to reduce the loss value. 
A variable $x$ is more important if including it reduces more loss.
So defined, a variable $x$ with $\mathcal{I}(x) = 0$ does not
contribute to determining the target $\bfrm[y]$, and can be neglected.
In the other extreme where $\mathcal{I}(x)=1$,  
the variable $x$ dominates the prediction for $\bfrm[y]$, and 
other variables can be neglected.
 
The final 
performance of the ensemble is then evaluated at some test data, 
and is measured by $R^2$, defined as the fraction of the 
variance of the target values explained, in regression problems, 
and by the correct-classification rate, $r$, 
in classification problems 
\citep[see, e.g.,][for a detailed description]{ChenYangyao:2020:Halo-structure}. 

The $R^2$ value satisfies the condition 
\begin{equation}
	0 \le R^2 \le 1.
\end{equation}
If a regressor has $R^2=1$, the relation between $\bfrm[x]$ and $\bfrm[y]$
is deterministic. On the other hand, $R^2=0$ indicates that there is no significant 
correlation between $\bfrm[x]$ and $\bfrm[y]$. Thus, $R^2$ measures the correlation 
strength between the predictor and target variables. 

As a simple demonstration, we consider an example where $x$ and $y$ satisfy 
a linear relation $y= k x+\epsilon$, with $\epsilon$ being a Gaussian random 
noise of zero mean and a constant variance. In such a case,
the Pearson correlation coefficient $\rho_{x,y}$, reflects the correlation 
strength between the variable pair. If one builds a linear least-square regression 
model and calculates the $R^2$ defined above, one gets $R^2=\rho_{x,y}^2$, 
i.e. $R^2$ is just the square of the correlation coefficient.

If the variable pair have a non-linear relation or they are in high dimensional space, 
the Pearson correlation coefficient is not so meaningful. In such cases, $R^2$ is a 
natural extension that has an interpretation similar to that in the linear case.

It is a common practice to adopt a threshold value to determine 
whether a correlation is strong or not. For example, if $R^2 < 0.5$ we may conclude 
that most of the driving factors are still missing in the model. On the other 
hand, if $R^2 > 0.5$, we may conclude that the main factors driving the target 
variable have already been included in the predictor set.

\bsp	
\label{lastpage}
\end{document}